\RequirePackage{fix-cm}

\documentclass[twocolumn,epjc3]{svjour3}  
\smartqed  
\RequirePackage{graphicx}

\usepackage{amsmath,amssymb}
\usepackage[switch]{lineno}
\usepackage{xspace}
\usepackage{graphicx}  

\newcommand{\Hbar}{$\overline{\mathrm{H}}$ }
\newcommand{\pbar}{$\overline{\mathrm{p}}$ }

\newcommand{\MCPTwo}{\ensuremath{\mathrm{MCP_{up}}}\xspace}
\newcommand{\MCPFive}{\ensuremath{\mathrm{MCP_{\overline H}}}\xspace}

\renewcommand{\rubric}{}
\renewcommand{\makeheadbox}{}

\journalname{Eur. Phys. J. C}
\begin{document}

\title{A determination of the backscattering probability of low-energy antiprotons
}


\author{K. Park\thanksref{e1,addr1}
        \and E. Perez\thanksref{e2,addr2} 
        \and P.~Adrich\thanksref{addr12}
        \and I.~Belosevic\thanksref{addr3}
        \and P.~Cladé\thanksref{addr8}
        \and M.~Chung\thanksref{addr4}
        \and P.~Comini\thanksref{addr3}
        \and P.~Crivelli\thanksref{addr5}
        \and P.~Debu\thanksref{addr3}
        \and A.~Douillet\thanksref{addr8,addr8a}
        \and S.~Geffroy\thanksref{addr6}
        \and S.~Guellati-Khelifa\thanksref{addr8, addr8b}
        \and P.~Guichard\thanksref{addr7}
        \and P.-A.~Hervieux\thanksref{addr7}
        \and L.~Hilico\thanksref{addr8, addr8a}
        \and P.~Indelicato\thanksref{addr8}
        \and S.~Jonsell\thanksref{addr9}
        \and J.-P.~Karr\thanksref{addr8, addr8a}
        \and B.~Kim\thanksref{addr17}
        \and S.~Kim\thanksref{addr1}
        \and E.-S.~Kim\thanksref{addr10}
        \and N.~Kuroda\thanksref{addr11}
        \and B.~Lee\thanksref{addr1}
        \and L.~Liszkay\thanksref{addr3}
        \and D.~Lunney\thanksref{addr6}
        \and G.~Manfredi\thanksref{addr7}
        \and B.~Mansoulié\thanksref{addr3}
        \and V.~Martimort\thanksref{addr8}
        \and M.~Matusiak\thanksref{addr12}
        \and V.~Nesvizhevsky\thanksref{addr13}
        \and F.~Nez\thanksref{addr8}
        \and N.~Paul\thanksref{addr8}
        \and P.~Pérez\thanksref{addr3,addr16}
        \and C.~Regenfus\thanksref{addr5}
        \and C.~Roumegou\thanksref{addr6}
        \and J.-Y.~Roussé\thanksref{addr3}
        \and F.~Schmidt-Kaler\thanksref{addr14}
        \and K.~Szymczyk\thanksref{addr12}
        \and T.~A.~Tanaka\thanksref{addr11, addr15}
        \and B.~Tuchming\thanksref{addr3}
        \and D.-P.~van~der~Werf\thanksref{addr16}
        \and D.~Won\thanksref{addr1}
        \and S.~Wronka\thanksref{addr12}
        \and P.~Yzombard\thanksref{addr8} \\
        (GBAR Collaboration)
}

\thankstext{e1}{e-mail: kwanhyung.park@cern.ch}
\thankstext{e2}{e-mail: emmanuel.perez@cern.ch}


\institute{Department of Physics and Astronomy, Seoul National University, Seoul, Korea \label{addr1}
           \and CERN, EP Department, 1 Esplanade des Particules, 1217 Meyrin, Switzerland \label{addr2}
           \and National Centre for Nuclear Research (NCBJ), ul. Andrzeja Soltana 7, 05-400 Otwock, Swierk, Poland \label{addr12}
           \and IRFU, CEA, Université Paris-Saclay, F-91191 Gif-sur-Yvette, France \label{addr3}
           \and Laboratoire Kastler Brossel, Sorbonne Université, CNRS, ENS-Université PSL, Collège de France, Campus Pierre et Marie Curie, 4, Place Jussieu, 75005, Paris, France \label{addr8}
           \and The Pohang University of Science and Technology (POSTECH), Pohang, Republic of Korea \label{addr4}
           \and Institute for Particle Physics and Astrophysics, ETH Zurich, 8093 Zurich, Switzerland \label{addr5}
           \and Université d’Evry-Val d’Essonne, Université Paris-Saclay, Boulevard François Mitterand, 91000 Evry, France \label{addr8a}
           \and Université Paris-Saclay, CNRS/IN2P3, IJCLab, Orsay, France \label{addr6}
           \and Conservatoire National des Arts et Métiers, 292 rue Saint Martin, 75003 Paris, France \label{addr8b}
           \and Université de Strasbourg, CNRS, IPCMS, UMR 7504, F-67000 Strasbourg, France \label{addr7}
           \and Department of Physics, Stockholm University, Stockholm, Sweden \label{addr9}
           \and Center for Underground Physics, Institute for Basic Science, Daejeon, Korea \label{addr17}
           \and Department of Accelerator Science, Korea University Sejong Campus, Sejong, Korea \label{addr10}
           \and Institute of Physics, University of Tokyo, Tokyo, Japan\label{addr11}
           \and Institut Max von Laue - Paul Langevin (ILL), Grenoble, France \label{addr13}
           \and Department of Physics, Swansea University, Swansea, United Kingdom \label{addr16}    
           \and QUANTUM, Institut für Physik, Johannes Gutenberg Universität, Mainz, Germany \label{addr14}
           \and \emph{Present Address:} National Metrology Institute of Japan (NMIJ), National Institute of Advanced Industrial Science and Technology (AIST), Tsukuba, Japan\label{addr15}        
}

\date{}

\maketitle
\thispagestyle{empty}

\begin{abstract}
It is commonly assumed that antiprotons impinging on a material surface annihilate promptly with the nuclei of the material. However, at kinetic energies of a few keV, this assumption may not hold.
As with low-energy protons, electrons or positrons that can be reflected from a target, they may undergo large-angle Coulomb scattering before annihilation occurs, thereby appearing to be ``backscattered" from the material surface.
This backscattering fraction, largely unknown, is a crucial ingredient to the determination of the production cross-section  of antihydrogen atoms in the GBAR experiment.
This paper presents a determination of the probability that 4 and 6~keV antiprotons backscatter on the surface of a Micro-Channel Plate detector used for beam imaging at GBAR.
No evidence for backscattering has been found and an upper limit of $14\%$ at $68\%$ confidence level has been set on this probability.
The impact of backscattering on the determination of the number of antiprotons that participate in antihydrogen production in GBAR is also addressed.

\end{abstract}

\section{Introduction}
\label{intro}

\begin{figure*}[tbh]
  \centering
    \includegraphics[width=0.9\linewidth]{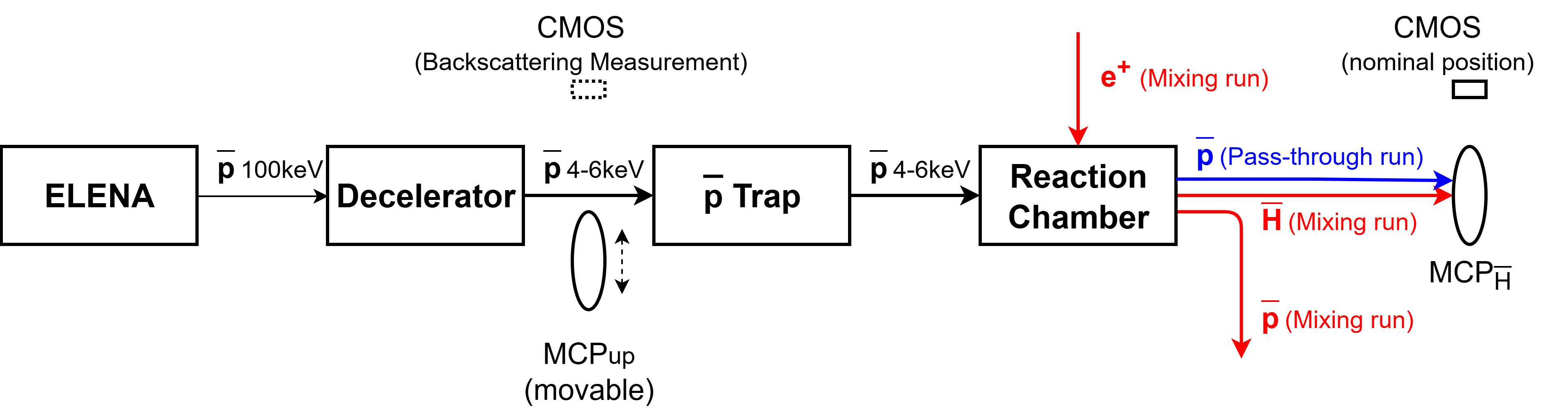}
    \caption{Simplified schematic of the GBAR antiproton beam line (not to scale). 
    In “pass-through” runs, positrons are not injected into the reaction chamber and the antiprotons reach \MCPFive (blue arrow). The configuration for standard mixing runs, aiming at antihydrogen production, is indicated by the red arrows:  a deflecting electrode prevents the antiprotons to reach the MCP and only the \Hbar atoms produced in the reaction chamber are detected on \MCPFive. Other movable MCPs can be inserted in the beam line to image the beam at various locations; only one of them, located upstream (\MCPTwo), is shown here. The CMOS sensor is drawn at its nominal position above \MCPFive; the dashed box above \MCPTwo marks its temporary location during the backscattering measurements.
    }
    \label{fig:BeamLine}
\end{figure*}

The GBAR\footnote{Gravitational Behaviour of Antimatter at Rest.} experiment~\cite{Adrich:2023tua} is operated at the Antiproton Decelerator – Extra Low ENergy Antiproton ring (AD-ELENA) facility at CERN, which delivers 100~keV antiprotons ($\overline{\mathrm{p}}$) in shots of typically $12 \times 10^6$ particles.
The primary objective of the 2024  run was to measure the antihydrogen 
($\overline{\mathrm{H}}$)
production cross-section through the charge exchange reaction of antiprotons, decelerated to 4 or 6~keV, with a positronium cloud. The measurement is described in Ref~\cite{GBAR-xsection}.
An essential quantity for this measurement is the  number of antiprotons passing through the cavity where the reaction takes place.
To quantify this antiproton flux, dedicated ``pass-through" runs are taken regularly, during which antiprotons traverse the cavity  and continue downstream to a Micro-Channel Plate (MCP) detector, called\footnote{In standard runs, an electrode deflects these antiprotons and only the neutral \Hbar atoms reach \MCPFive on which they annihilate. The \MCPFive being the device that detects the produced \Hbar atoms, it is called ``the detection MCP'' in Ref.~\cite{GBAR-xsection}.} \MCPFive. When they annihilate on \MCPFive, secondary particles from these annihilations (predominantly charged pions) reach a small Complementary Metal-Oxide-Semiconductor (CMOS) detector, creating pixel clusters in the sensor. The sensor is installed above the beam line at \MCPFive, outside the vacuum chamber, facing the MCP. With appropriate calibration, the number of observed clusters can be converted to the number of incident antiprotons. This method is detailed in Ref.~\cite{CR_paper}, where more information about the CMOS detector can be found. A simplified sketch of the GBAR antiproton line is shown in Fig.~\ref{fig:BeamLine}.

A fraction of the 4 or 6~keV antiprotons that reach  \MCPFive may not annihilate on the MCP but undergo large angle Coulomb scattering before annihilation occurs, thereby appearing to be  ``backscattered" from the MCP surface.
While the backscattering of low-energy electrons, positrons and protons has been extensively studied experimentally and is routinely modelled using Monte-Carlo transport codes such as SRIM/TRIM~\cite{Ziegler:2010bzy}, the physical picture for antiprotons is qualitatively different. Electrons and positrons are scattered predominantly by Coulomb interactions and lose energy continuously in the material, while protons may additionally undergo elastic nuclear scattering and emerge either as protons or, following electron capture near the surface, as neutral hydrogen atoms. In contrast, antiprotons compete with atomic capture and subsequent annihilation. Their backscattering probability therefore reflects the competition between Coulomb scattering, energy loss and annihilation, and cannot be inferred directly from measurements with ordinary charged particles. 
A non-vanishing backscattering fraction may affect the aforementioned calibration of the CMOS detector (namely the relation between the number of clusters seen in the sensor and the number of antiprotons reaching the MCP), since the CMOS detector may be blind (or partially blind)  to antiprotons that backscatter off the MCP.

In standard ``mixing" runs, the same MCP is used to count the number of produced \Hbar atoms, that fly straight to it (see Fig.~\ref{fig:BeamLine}).  The anti-atom is stripped off from its positron in the first atomic layers of the material and the annihilation of the antiproton produces a signal in the MCP - unless the \pbar backscatters, with the same probability as for a bare antiproton. Backscattering would then lead to an apparent inefficiency that should be corrected for in the determination of the \Hbar production cross-section.

The probability for such antiproton backscattering to occur is  largely unknown. Some evidence for a rather large backscattering fraction was reported in~\cite{Bianconi:2008up,Bianconi:2009bsc}, for a non monochromatic beam (down to 1 keV) hitting an aluminium wall. While this observation may be indicative of a sizable probability for 4 keV or 6 keV antiprotons to backscatter off the MCP used for the GBAR measurements, it can not be used to provide a quantitative estimation for the latter. This paper shows how this backscattering fraction has been determined in-situ, using dedicated measurements in the CMOS sensor together with some input from a GEANT4~\cite{GEANT4:2002zbu} simulation. The structure is as follows.
Section 2 presents GEANT4 predictions for this fraction and introduces the measurement method: the dependence of the probability to see a signal in the CMOS sensor with the distance between the sensor and the MCP differs for antiprotons that annihilate on the MCP and for those that backscatter; measuring the variation of the signal with this distance thus provides a handle on the backscattering fraction. The experimental setup is described in Section 3, together with its simulation. The measurements are presented in Section 4, and the extraction of the backscattering fraction is made in Section 5. The calibration factor of the CMOS detector, required to determine the number of incident antiprotons from the pass-through runs, and its sensitivity to backscattering, are obtained in Section 6. More details can be found in~\cite{GBARNote,KwanHyungThesis}.

\section{GEANT4 predictions of antiproton backscattering}

Simulations based on GEANT4 (version 11.3) have been conducted in order to quantify the fraction of low energy antiprotons that backscatter off a surface, especially when they hit this surface at normal incidence. Several surface materials have been considered: chro\-mium and nickel, of which the metallic coating layer of the MCPs used in GBAR is made, but also others, like gold, as an example of a much heavier material. 

\subsection{Modelling of antiproton annihilation}\label{sec:modelling} 
All simulations presented here make use of the {\tt{INCLXX}} physics list of GEANT4. As shown in Ref.~\cite{CR_paper}, the modelling of antiproton annihilation in this physics list reproduces the multiplicity of charged pions measured in such annihilations and its dependence with the atomic mass of the target material, in contrast to the {\tt{BERT}} and {\tt{BIC}} lists. However, the angular distribution of the particles created in these annihilations show large, unphysical anisotropies.
An ad-hoc procedure has been used to recover isotropic angular distributions, whereby the direction of the momentum of particles created in the annihilation process is randomised~\cite{KwanHyungThesis}.

\subsection{Electromagnetic physics list}
\label{sec:G4_EMlist}

The Coulomb interactions experienced by an antiproton penetrating a surface have been simulated using different options offered by GEANT4:
\begin{itemize}
    \item a ``single scattering'' (SS) model, 
    in which GEANT4 generates individual Coulomb interactions - leading to a very large number of steps along the antiproton track;
    \item a ``multiple scattering'' (MSC) model which averages the effects of individual Coulomb interactions. Compared to the SS model, this averaging results in a drastic speed-up of the simulation. The so-called ``WentzelVI'' model is used.
    Two configuration parameters have a crucial impact on the predicted fraction of backscattering:
    \begin{itemize}
        \item The WentzelVI model actually allows for a combination of single scattering and multiple scattering simulations:
        large-angle scattering, with a scattering angle larger than a $\theta_{lim}$ parameter, are modelled with SS, while MSC is used for scattering angles below this limit. In the default configuration, $\theta_{lim}$ is set to 180 degrees, which means that the SS model is never used. This default value is unlikely to be a good choice for predicting the backscattering of antiprotons, for which large- or medium-angle scatterings are expected to play a significant role.
        \item a dimensionless ``range cut'' parameter that limits the size of the GEANT4 steps. 
        The value of this parameter, in the default configuration of the  WentzelVI model, is actually rather large for antiprotons (0.2, while it is $0.04$ for electrons). With such a large value, antiprotons can go deeply into the volume in the first step of the tracking; after which they may have lost too much energy to escape, and the backscattering fraction can be underestimated.
    \end{itemize}    
\end{itemize}

When the $\theta_{lim}$  parameter is smaller than about 2 radians, the backscattering probability predicted by GE\-ANT4 for 6~keV antiprotons that hit a 1~mm thick stub of gold shows little sensitivity to the values of $\theta_{lim}$ and of the range cut. For larger values of $\theta_{lim}$, the prediction critically depends on the range cut: for small values of the latter (a few $10^{-2}$), the low $\theta_{lim}$ prediction is recovered, but it decreases by a factor of 2 when the range cut increases to its default value of 0.2 while $\theta_{lim}$ is set to 2.5~radians. This variation reaches a factor of about 5 in the extreme case where $\theta_{lim}$ is set to $180^{\circ}$. \\

Unless explicitly stated otherwise, all subsequent simulations have been made with the WentzelVI  model, with $\theta_{lim}$ set to 0.2~radians and the range cut to its default value. This choice was made because it leads to a backscattering fraction that is very close to that predicted by the SS model (which is a priori expected to be more accurate), for a simulation time that remains affordable. An alternative choice will be considered for assessing systematic uncertainties.

\begin{figure}[tbh]
  \centering
    \includegraphics[width=1.\linewidth]{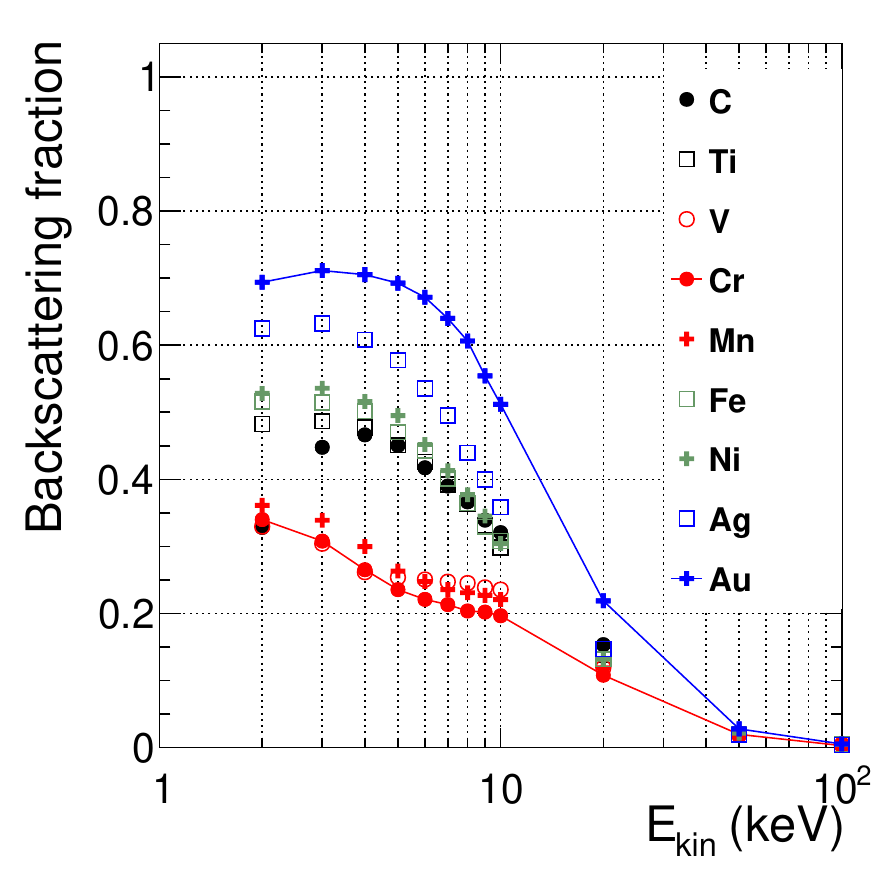}
    \caption{GEANT4 predictions for the backscattering fraction of antiprotons that  hit a 20~nm thick stub of material at normal incidence, as a function of their kinetic energy,  for several examples of the stub material.  }
    \label{fig:G4_backscattering_vs_Ekin_various_materials}
\end{figure}

\subsection{Impact of the material}

Figure~\ref{fig:G4_backscattering_vs_Ekin_various_materials} shows the probability that antiprotons backscatter off a 20 nm thick stub of a given material, as a function of their kinetic energy, for various elements ranging from very light, such as carbon, to rather heavy atoms like gold. Note that, in order to predict the probability that an antiproton backscatters off a MCP detector, the numbers resulting from this simulation should be rescaled according to the open area ratio of the detector, namely the percentage of the total effective area that is made up of open channels. For example, for a MCP whose coating is made of chromium and which has an open ratio of $63 \%$, the probability that 6~keV antiprotons backscatter is predicted to be $23 \% \times ( 1 - 0.63) = 8.5 \%$.

In Fig.~\ref{fig:G4_backscattering_vs_Ekin_various_materials}, 
it can be seen 
that the backscattering fraction depends significantly on the material,
which is a priori expected. However, since the detailed electronic or shell structure of the elements is unknown to the multiple scattering model of GEANT4, the $Z$-dependence of this predicted backscattering fraction calls for questions. The\-re is no reason why the prediction should be so much lower for V, Cr and Mn ($Z = 23, 24$ and $25$, respectively) than for Ti ($Z = 22$) or Fe ($Z = 26$). The cause for this behaviour  has been tracked down to the stopping power of antiprotons~\cite{GBARNote}, as implemented in this  version of GEANT4. It is unexpectedly large for V, Cr and Mn, and generally, the agreement with experimental measurements is quite worse than what  
was reported in Ref.~\cite{Chauvie:2007zz}. \\

Due to the large model dependence of the predicted backscattering fraction, and to the 
questionable 
\pbar stopping power that is used in this version of GEANT4, the GEANT4 predictions for the backscattering probability of low energy antiprotons are very uncertain, and an experimental measurement  would be most welcome. The next section shows that an experimental determination can be made, by measuring how the signal seen in the CMOS sensor varies with the distance between the sensor and the centre of the MCP. Indeed, the GEANT4 predictions for this dependence are seen to be much more robust than those for the absolute backscattering fraction.

\begin{figure*}[tbh]
  \centering
  \begin{tabular}{cc}
    \includegraphics[width=0.45\linewidth]{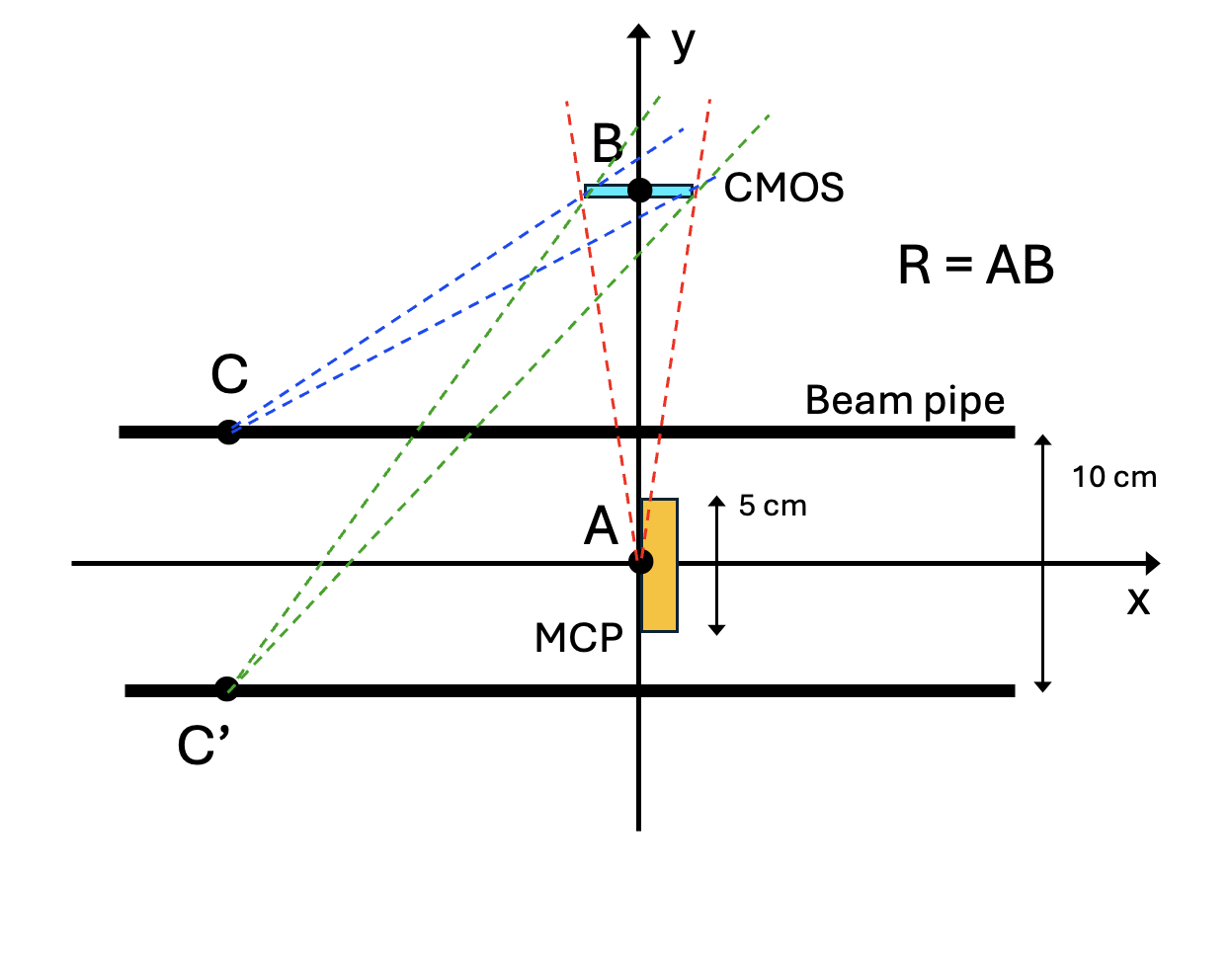} &
    \includegraphics[width=0.45\linewidth] {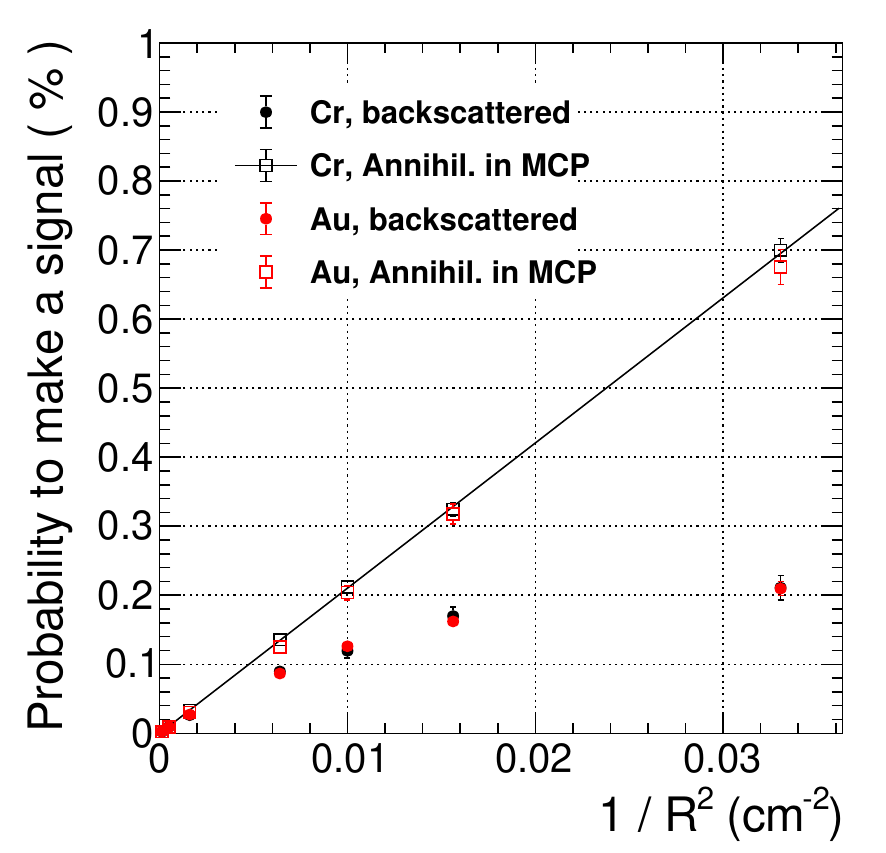}
    \end{tabular}
    \caption{Left: Sketch of the simplified model used in section~\ref{sec:bs_simplified_model}. The incoming antiprotons travel along the $x$-axis and hit the MCP in its centre (A). The point B corresponds to the centre of the CMOS sensor. Points C and C' show example positions at which antiprotons  backscattering off the MCP could annihilate. The cones spanned by the dashed lines indicate the geometric acceptance of the CMOS sensor for antiprotons annihilating in A, C or C'. Right: Probability that a charged particle crosses the CMOS sensor as a function of the inverse of the square of the  distance $R$ between the sensor and the centre of the MCP, obtained from a GEANT4 simulation using this simplified model. The rightmost point corresponds to $R = 5.5$~cm. The open squares (closed circles)  show this probability for antiprotons that annihilate on the MCP (backscatter). Black (red) symbols correspond to the configuration where the MCP coating is made of chromium (gold). The  line shows the result of a linear fit to the open black squares.}
    \label{fig:G4_simplified_proba_vs_distance}
\end{figure*}

\subsection{Predicted signal in the CMOS sensor: simplified model}
\label{sec:bs_simplified_model}

With an isotropic angular distribution for the particles created when antiprotons hit the MCP in its centre and annihilate, the CMOS sensor would naively intercept a fraction $S / 4 \pi R^2$ of the annihilation products, where $S$ denotes the small active area of the sensor (about $60$~mm$^2$), and $R$ the distance between the centre of the CMOS sensor and the centre of the MCP. \\
The  dependence of the signal expected in the CMOS sensor, as a function of the distance between the sensor and the centre of the MCP, has been  studied first with a most simplified model, 
illustrated in the left panel of Fig.~\ref{fig:G4_simplified_proba_vs_distance}. 
Antiprotons, travelling along the $x$ direction inside a stainless-steel beam pipe of internal radius 5~cm, are intercepted by a MCP, modelled as a simple 2~mm thick disk of radius 2.5~cm, made of lead-glass, with a 20~nm coating layer in front of it. The front face of the MCP is in the plane $x=0$ and its centre is along the beam line.
A scoring surface centred around $x=0$, of the same dimensions as the CMOS sensor, is used to determine the probability that the sensor be crossed by a charged particle, separately for events where the antiproton annihilates on the MCP (on its coating or in its substrate\footnote{For 4~keV or 6~keV antiprotons, the annihilation usually happens inside the 20~nm coating layer, while 100 keV antiprotons  penetrate  deeper and annihilate inside the MCP substrate.}), and for events where it backscatters. 
Events have been simulated for several values of the  distance $R$, using two materials for the coating of the MCP, chromium and gold.
Figure~\ref{fig:G4_simplified_proba_vs_distance} (right) summarises the results for 6~keV antiprotons. Several  observations can be made from this figure:
\begin{itemize}
    \item despite the fact that GEANT4 predicts a very different backscattering fraction when 6~keV antiprotons are sent on a surface made of chromium or of gold (see Fig.~\ref{fig:G4_backscattering_vs_Ekin_various_materials}), the probabilities to observe a signal in the scoring surface, for antiprotons that annihilate and for those that backscatter, are  independent of the coating material;
    \item for antiprotons that annihilate on the MCP, the probability to see a signal grows linearly with $ 1 / R^2$, as expected;
    \item at large distances, the probabilities to see a signal, for antiprotons that annihilate and for those that backscatter, are very similar;
    \item at small distances, these two probabilities are different; it is less likely to see a signal for antiprotons that backscatter than for antiprotons that annihilate in the MCP.
\end{itemize}

\begin{figure*}[tbhp]
  \centering
    \centering
    \begin{tabular}{cc}
    \includegraphics[width=0.45\linewidth]{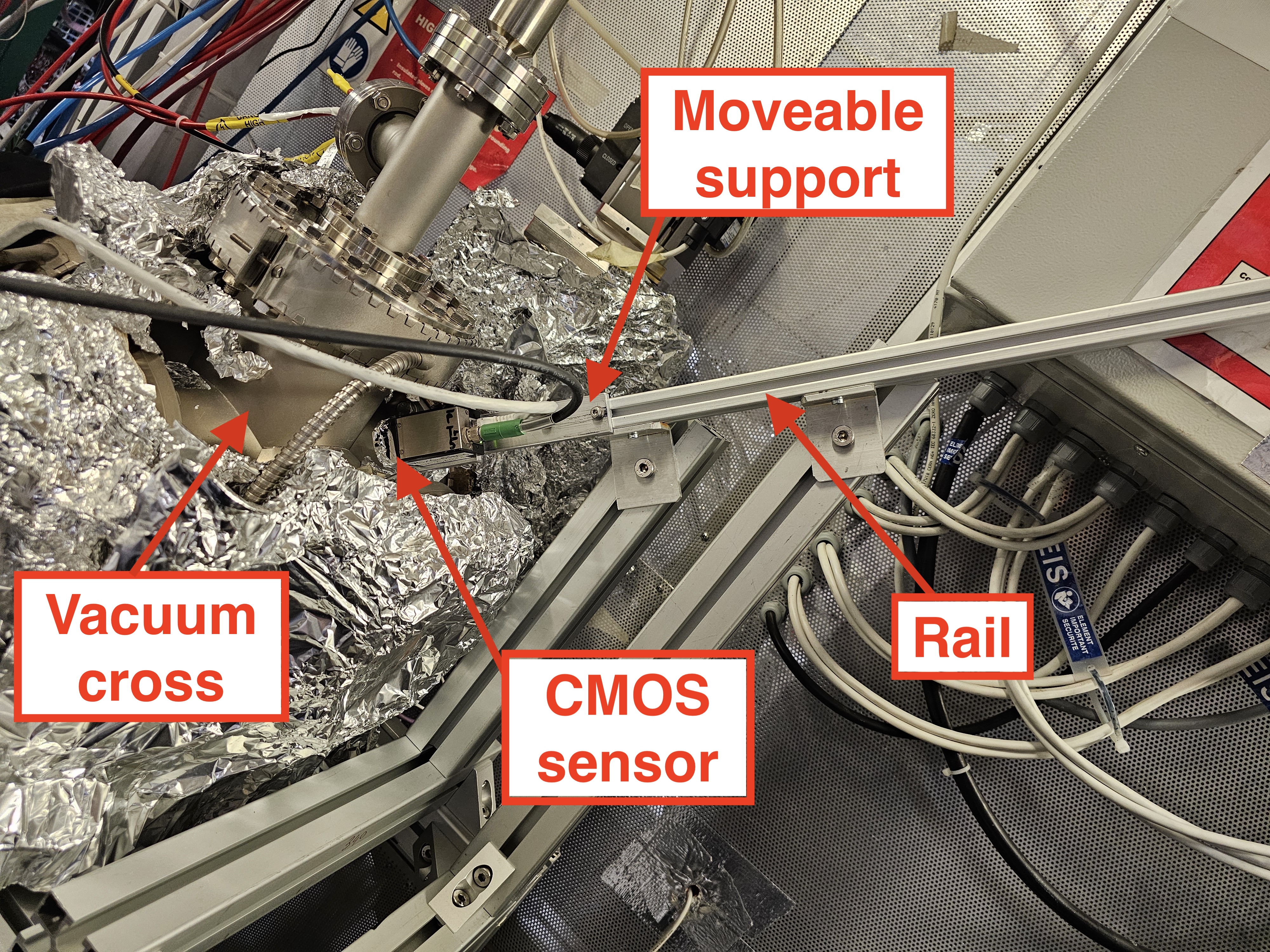} &
    \includegraphics[width=0.45\linewidth]{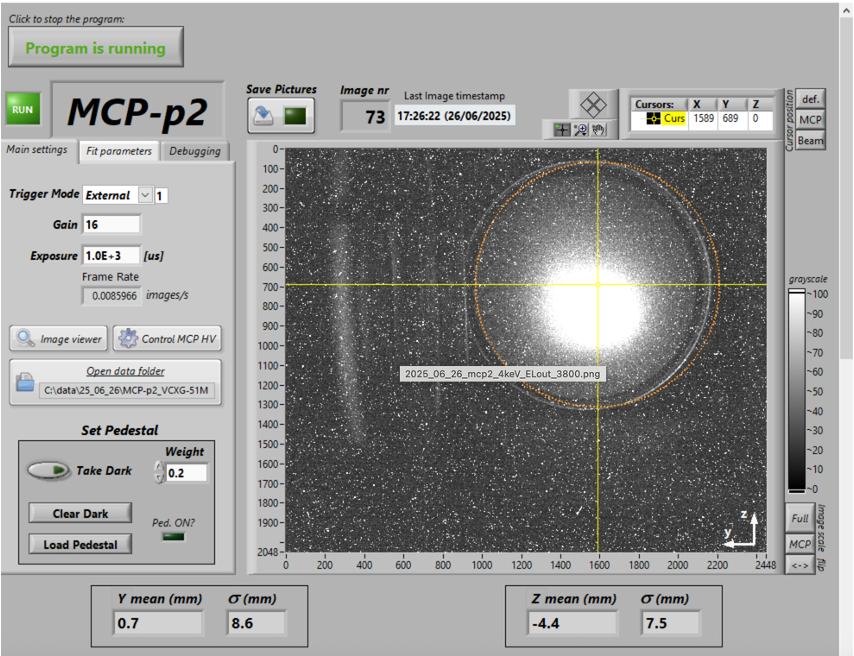}
    \end{tabular}
    \caption{Left: Mechanical setup used to measure the backscattering fraction on \MCPTwo, from the dependence of the signal observed in the CMOS sensor with the distance between the sensor and the centre of the MCP. The CMOS sensor, shown here at its closest position to the beam line, was manually moved along the rail seen in the picture. Right: Image of the spot of the 4 keV beam on \MCPTwo.}
    \label{fig:mcp2-setup}
\end{figure*}

When antiprotons backscatter off the MCP, they end up annihilating on the beam pipe, upstream of the MCP, on average by about 9~cm in this simplified model.   
The distribution of the longitudinal position (along the $x$-axis) where this annihilation happens depends on the angular distribution of the backscattered antiprotons. While the GEANT4 predictions for the backscattering fraction of antiprotons are significantly model- or assumption-dependent as was seen previously, the modelling of this angular distribution appears to be  more robust. 
When the distance between the CMOS sensor and the centre of the MCP is large ($R \gg 9$~cm), 
the geometric acceptance of the CMOS sensor is similar 
for antiprotons that annihilate on the MCP and for antiprotons that annihilate in the beam pipe. Consequently, at large distance, the probability to leave a signal is about the same\footnote{This fact means that backscattering should be irrelevant in the CMOS detector when the sensor is at a large enough distance from the MCP: the CMOS sensor will see the same signal, irrespective of the backscattering fraction. While Fig.~\ref{fig:G4_simplified_proba_vs_distance} may indicate that this asymptotic regime is reached already at a distance of the order of 40~cm, the next section will show that, depending on the exact geometry of the environment in the vicinity of the MCP, the two probabilities can still be significantly different at this distance. } for backscattered antiprotons and for antiprotons that annihilate on the MCP, as seen indeed in Fig.~\ref{fig:G4_simplified_proba_vs_distance}.
The difference between the dependence with $1/R^2$ of the two probabilities shown in Fig.~\ref{fig:G4_simplified_proba_vs_distance} provides a handle on the backscattering fraction. Assuming that the probability $P_{\rm{annihil}}(R)$ to see a signal for antiprotons that annihilate on the MCP, and the probability $P_{\rm{back}}(R)$ to see a signal for antiprotons that backscatter, are reliably predicted by the simulation (up to an overall normalisation factor), measuring the variation of the signal seen in the CMOS sensor in a large range of distances between the sensor and the centre of the MCP allows the backscattering fraction to be extracted. 
As shown in Fig.~\ref{fig:G4_simplified_proba_vs_distance}, $P_{\rm{annihil}}$ and $P_{\rm{back}}$ do not depend significantly on the material of the MCP coating. 
The robustness of these probabilities with respect to modelling assumptions has been investigated by varying the ``range cut"  and the $\theta_{lim}$ parameter. When the determination of $P_{\rm{back}}(R)$ is restricted to events where the backscattered antiproton does not re-scatter on the beam pipe, it shows little sensitivity to these variations, in contrast to the predicted backscattering fraction. Indeed, the prediction of $P_{\rm{back}}(R)$ then  depends primarily on the angular distribution of the backscattered antiprotons (which determines where they hit the beam pipe), and not on the modelling of the transport of the antiprotons in the MCP coating or material. Hence, it is largely insensitive to the stopping power in the material and to the detailed modelling of multiple scattering. The residual model dependence induced by potential re-scattering of the backscattered antiprotons on the beam pipe will be considered as a systematic uncertainty in Section~\ref{sec:backscattering-results}. 
In addition, a systematic uncertainty on the longitudinal position of the sensor with respect to the MCP was seen to have a significant impact on the prediction of $P_{\rm{annihil}}(R)$. 
This aspect will be illustrated in the next section, since the changes are even more significant with a more realistic simulation model.

\section{Experimental setup and its corresponding simulation}

\subsection{Experimental setup}

Following the study described in the previous section, dedicated data were taken in order to extract the \pbar  backscattering fraction from the dependence of the signal observed in the CMOS sensor with the  distance between the sensor and the centre of the MCP. 
The antiproton beam was intercepted by one of the movable MCPs installed in the GBAR line, called \MCPTwo (see Fig.~\ref{fig:BeamLine}), located immediately downstream of the GBAR decelerator~\cite{Adrich:2023tua}.
The CMOS sensor, which, during standard data taking,  is used to determine the flux of antiprotons that enter GBAR's reaction chamber, was temporarily removed from its nominal position and  installed above the beam line, at the \MCPTwo location. This location was chosen because it offers more flexibility to position the CMOS detector. 
The CMOS detector was fixed at different  positions by sliding it along a rail, pointed orthogonally to the beam line axis. The corresponding mechanical setup is shown in Fig.~\ref{fig:mcp2-setup} left.

\begin{figure*}[tbhp]
    \centering
    \begin{tabular}{cc}
    \reflectbox{\includegraphics[width=0.35\textwidth]{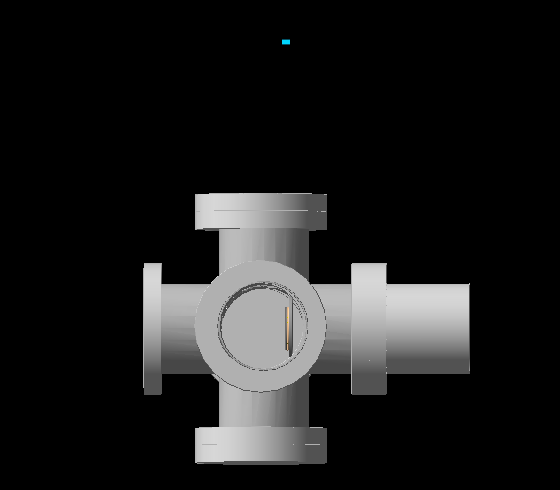}} &  
    \includegraphics[width=0.62\linewidth]{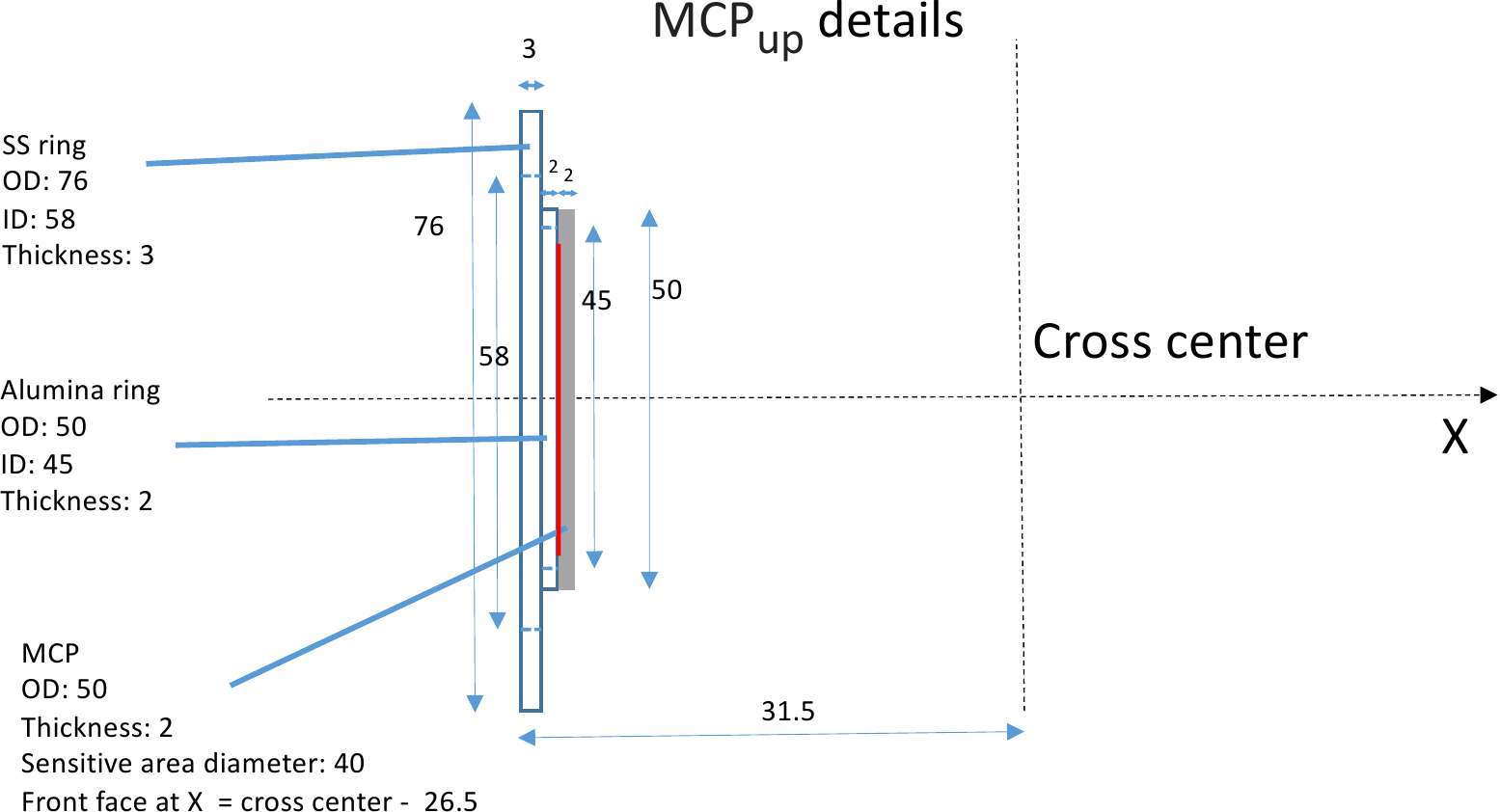} 
\end{tabular}
  \caption{Left:Visualisation of the simulation geometry around \MCPTwo. For visualisation purposes, the blank flange on the viewer side was removed to expose the internal components. The small light blue cuboid above is the CMOS detector. The antiprotons travel along the horizontal direction and arrive from the left side of the figure. Right: Schematic drawing of the \MCPTwo and nearby components. The beam travels along the $x$-axis.}
  \label{fig:MCP2}
\end{figure*}

The data have been collected in two periods of a few days each: in May 2025, at beam energies of 6~keV and 100~keV, and in June 2025, at beam energies of 4~keV and 100~keV. The CMOS sensor was set at distances of 9.3~cm, 11.3~cm, 14.3~cm, 19.3~cm, 39.3~cm, 84.3~cm and 94.3~cm. The mechanical setup had been moved between May and June, hence its position was not exactly the same in the two campaigns of data taking. 
An example of the image of the 4~keV beam on \MCPTwo is shown in Fig.~\ref{fig:mcp2-setup} right.

\subsection{Actual GEANT4 model}
\label{sec:G4_fullmodel}

In order to measure the \pbar backscattering fraction  using the method described in section~\ref{sec:bs_simplified_model}, the probabilities $P_{\rm{annihil}}( R )$ and $P_{\rm{back}}( R ) $ must be determined with a simulation model that reflects closely the experimental environment. To this end, a detailed modelling of the beam line components at the exit of the decelerator and of the \MCPTwo itself has been implemented in GEANT4.

The simulation models the beam line geometry using a vacuum 6-way cross with four blank flanges and one beam pipe positioned upstream of the \MCPTwo, as shown in the left panel of Fig.~\ref{fig:MCP2}. All beam line components 
are modelled as stainless-steel. The MCP substrate and its coating are modelled as plain disks. The substrate consists of a commonly used lead-glass\footnote{The so-called Corning 8161 composition is used~\cite{wiza}, with about $48 \%$ of lead, $26 \%$ of oxygen and $18 \%$ of silicon.}. 
The coating consists of a 20 nm thick layer of chromium.  
The MCP is mechanically supported by mounting rings positioned in front of the detector surface. Details of the implemented geometry are shown in the right panel of Fig.~\ref{fig:MCP2}. The simulation shows that these rings, which are structural components of the MCP assembly, have a significant impact on the  probabilities $P_{\rm{annihil}}( R )$ and $P_{\rm{back}}( R ) $, as will be seen below. 
 The CMOS detector is modelled as a $8.445 \times 7.065$~mm$^2$ rectangular silicon slab with a thickness of 10~$\mu$m. No internal structure is included in the model; the entire volume is treated as sensitive detector material.  
 It has been checked that the efficiency of the CMOS detector does not depend on the incidence angle of the impinging charged particle, for the range of angles relevant for this analysis~\cite{CR_paper}. \\

 \begin{figure}[tbhp]
  \centering
    \includegraphics[width=1.0\linewidth]{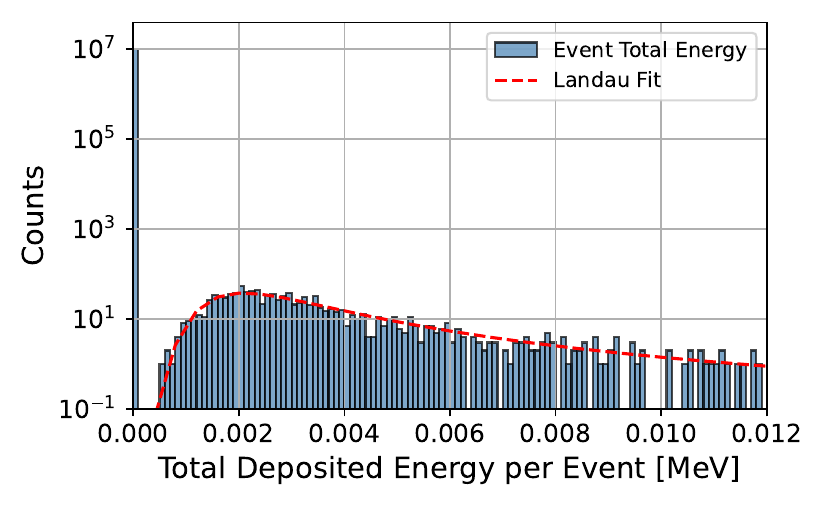}
    \caption{Energy deposited in the CMOS detector per event, as predicted from the simulation, when the CMOS sensor is at a radial distance of   
    $\sim 40$~cm   
    from the centre of \MCPTwo. In total, $10^{7}$ antiproton annihilations have been simulated, of which 1214 events lead to a non-zero energy deposit in the sensor. The peak at zero, which contains the vast majority of the events, corresponds to events for which no charged particle reaches the CMOS sensor, i.e. it reflects the geometric acceptance of the CMOS detector.  
    }
    \label{fig:MCP2_result_event}
\end{figure}

\begin{figure*}[tbhp]
  \centering
   \begin{tabular}{cc}
    \includegraphics[width=0.45\linewidth]{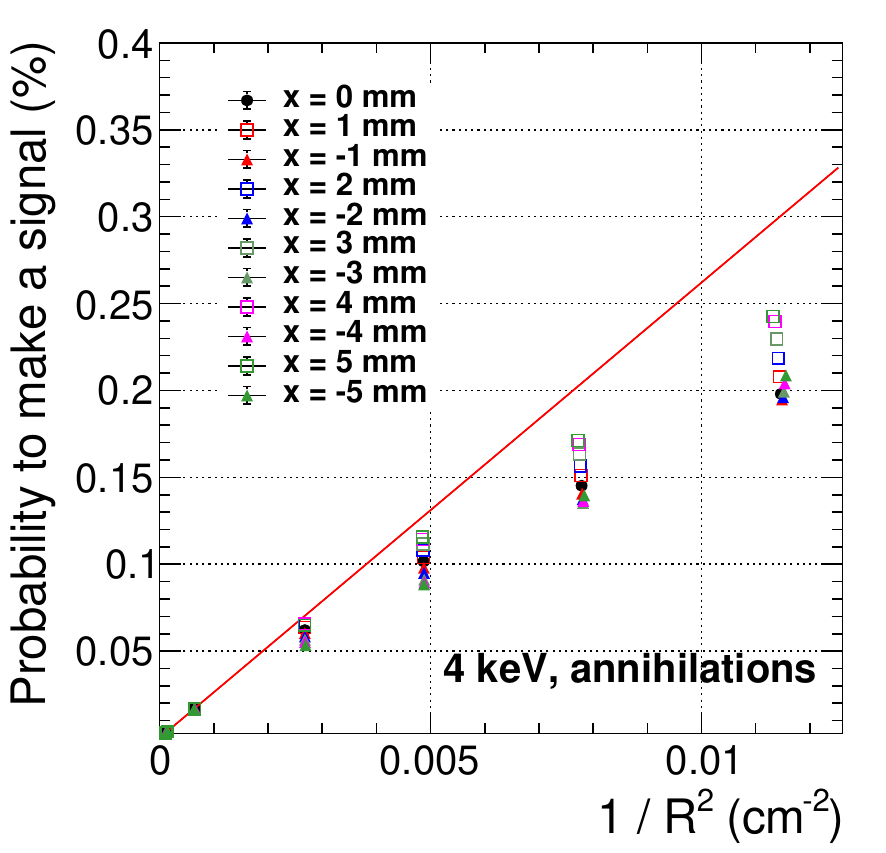}   & 
    \includegraphics[width=0.45\linewidth]{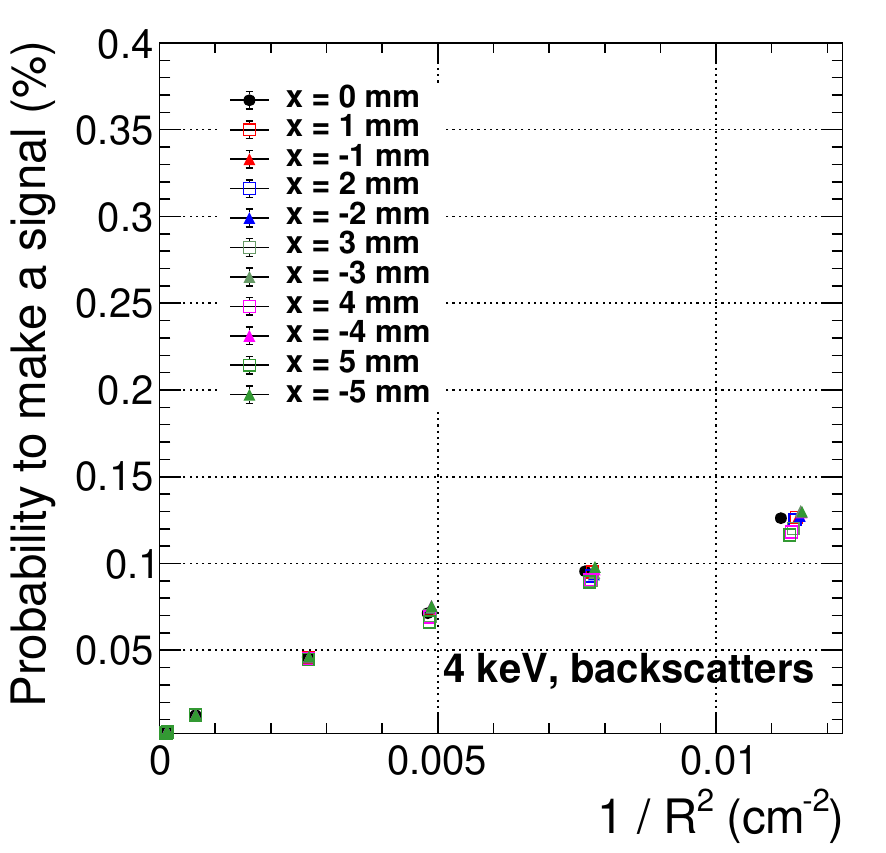}
  \end{tabular}
   \caption{Probability that a charged particle crosses the CMOS sensor when a 4 keV antiproton beam is sent onto \MCPTwo, as predicted from the GEANT4 simulation, as a function of the inverse of the squared distance between the sensor and the centre of the MCP. The probabilities are shown separately for (left) antiprotons that annihilate on the MCP  and for (right) those that backscatter, and for several offsets of the sensor along the longitudinal direction. By convention, the reference $x=0$ corresponds to the experimental setup, where the CMOS sensor was installed 8.5~mm upstream of the MCP.  In the left plot, the red line shows the result of a linear fit to the points at $x=0$ and $R \geq 39$~cm.}
    \label{fig:G4_final_templates_4keV}
\end{figure*}

Events have been simulated for kinetic energies of 100~keV, 6~keV and 4~keV and for the  positions of the CMOS sensor for which measurements have been made. 
The incoming antiprotons travel along the nominal beam direction, according to a distribution that reproduces the beam spot observed on \MCPTwo, at each energy. The 
beam shape was seen to have a negligible effect, but the position at which the beam centre hits the MCP is relevant. 
The statistical uncertainty on the predicted probability $P_{\rm{annihil}}( R )$ is below one percent in the whole range of $R$ considered. For $P_{\rm{back}}( R )$, it remains below $2 \%$ at 4~keV and 6~keV.
These probabilities are defined as the fractions of events for which energy is deposited in the CMOS sensor. An example distribution of this deposited energy is shown in Fig.~\ref{fig:MCP2_result_event}, which exhibits a characteristic Landau shape.  \\

Figure~\ref{fig:G4_final_templates_4keV} shows the probabilities $P_{\rm{annihil}}$ and $P_{\rm{back}}$ that a signal be observed in the CMOS sensor when a 4~keV antiproton beam is sent onto \MCPTwo, as predicted from the GEANT4 simulation, as a function of the inverse of the squared  distance between the sensor and the MCP centre. They are shown for several offsets of the sensor along the longitudinal direction, in a range of $\pm 5$~mm around the nominal position
of the CMOS sensor in the experimental setup (8.5~mm upstream of the MCP). Similar probabilities have been computed for the 6~keV and the 100~keV beams. In the left plot, a linear fit is shown to the points corresponding to the nominal longitudinal position of the sensor (labelled $x=0$) and to  $R \geq 39$~cm. It is clearly seen that the extrapolation of this fit to small  distances does not describe the GEANT4 results. This deviation from the naively expected linear behaviour (that was observed in Fig.~\ref{fig:G4_simplified_proba_vs_distance}) was traced to be due to material effects, in particular to additional pion absorption by the stainless steel MCP support ring~\cite{GBARNote}. 
The strong dependence of $P_{\rm{annihil}}$ with the longitudinal position of the sensor at small values of $R$, that is seen in Fig.~\ref{fig:G4_final_templates_4keV} left, has the same origin.

\section{The measurements on \MCPTwo}
\label{sec:MCP2-measurements}

\begin{figure*}[htbp]
  \centering
  \begin{tabular}{cc}
    \includegraphics[width=0.4\linewidth]{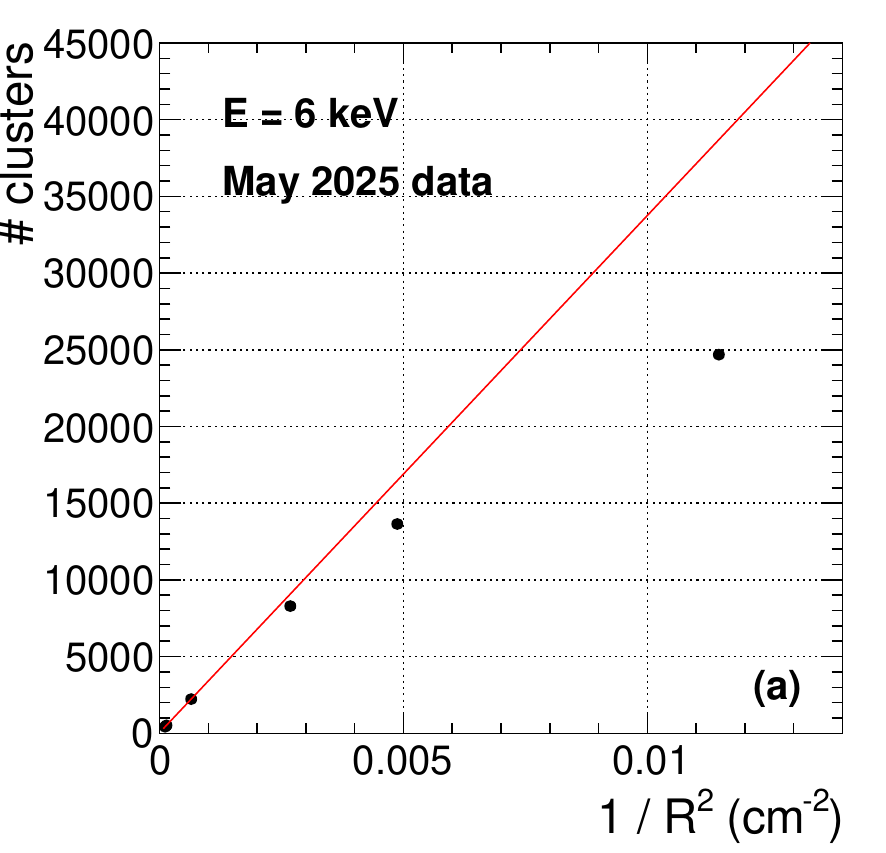} &
    \includegraphics[width=0.4\linewidth]{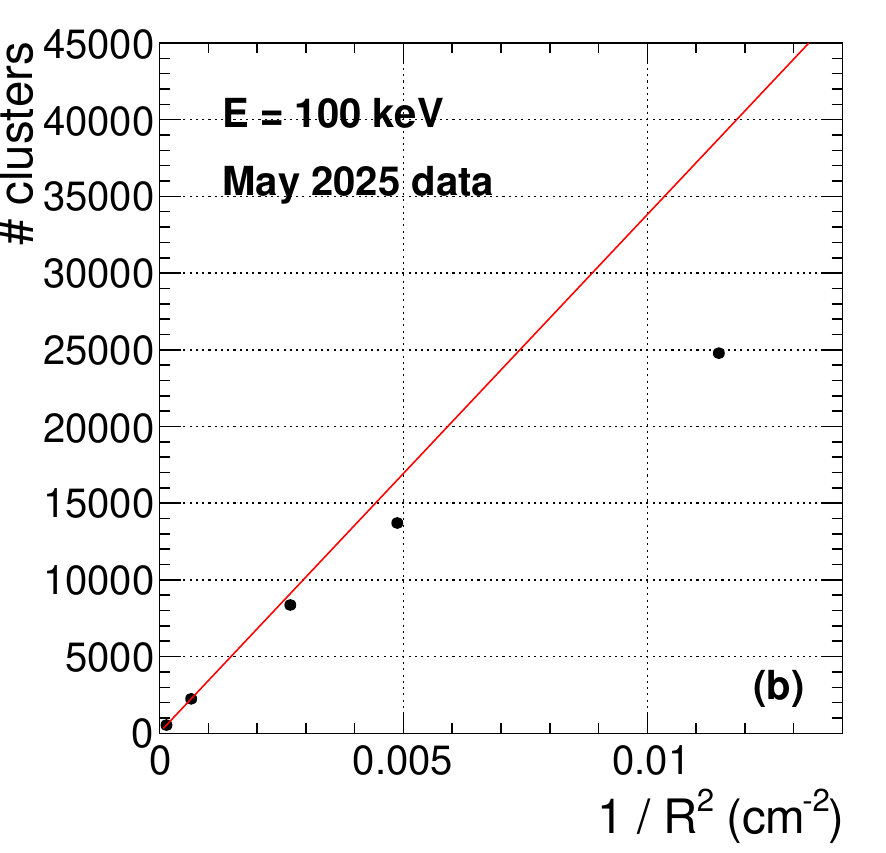} \\
    \includegraphics[width=0.4\linewidth]{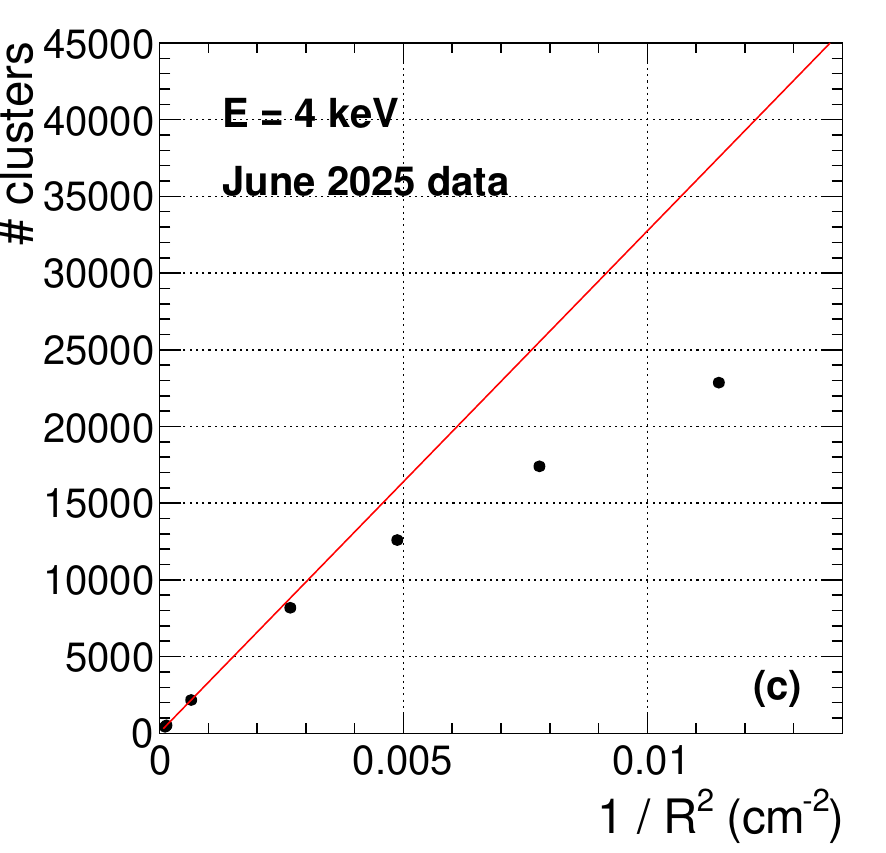} &
    \includegraphics[width=0.4\linewidth]{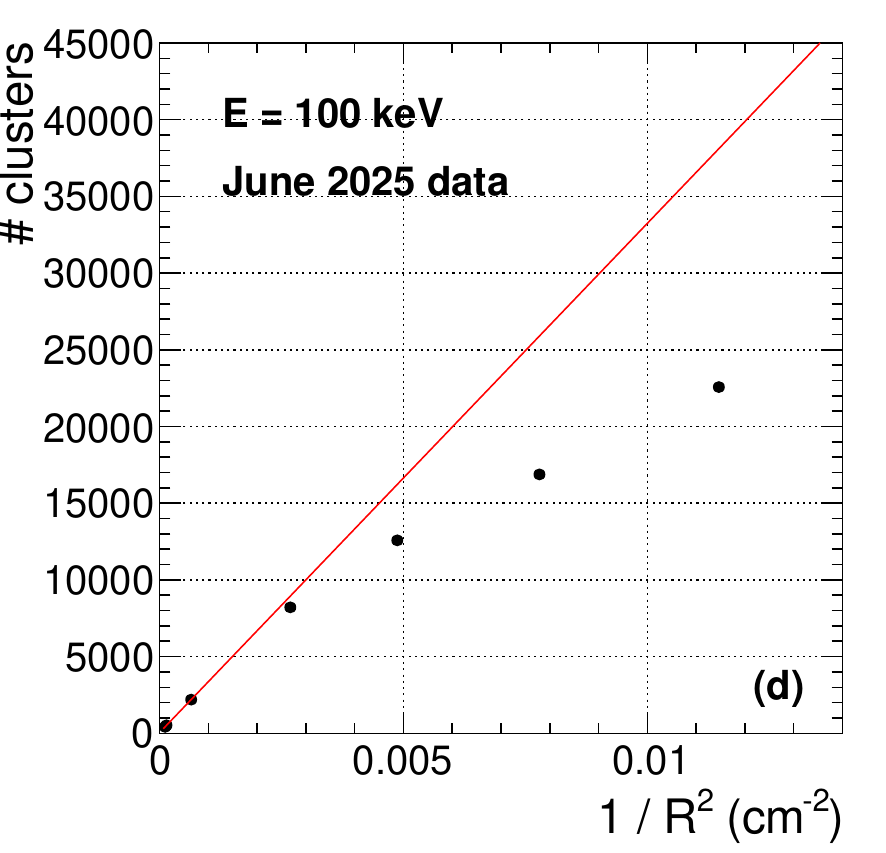}
    \end{tabular}
    \caption{Average number of clusters measured in the CMOS sensor as a function of the inverse of the squared  distance between the CMOS sensor and the centre of the MCP (data points), for (a) the 6~keV  and (b) the 100~keV data taken in May 2025, and for (c) the 4~keV and (d) the 100~keV data taken in June 2025. In each panel, the red line shows the result of a linear fit to the large distance measurements ($R \geq 39$~cm). }
    \label{fig:mcp2-Nclusters-raw-data}
\end{figure*}

A simple clustering algorithm is used to process the ``hits'' observed in the CMOS sensor. Only the pixels for which the deposited charge exceeds 200 Analog-to-Digital Converter (ADC) counts are considered - this threshold comes from an analysis of the pedestals, similar to that reported in Ref.~\cite{CR_paper}. Contiguous pixels with a charge above this threshold are clustered together, and clusters with a total charge above 500 ADC counts are selected. 
The  average number of clusters measured in all runs is shown in Fig.~\ref{fig:mcp2-Nclusters-raw-data}. Comparing Fig.~\ref{fig:mcp2-Nclusters-raw-data}b and Fig.~\ref{fig:mcp2-Nclusters-raw-data}d, a small difference is seen between the two sets of measurements taken at 100~keV, which reflects the fact that the position of the CMOS sensor is not exactly the same in the two campaigns.
Moreover, within a given campaign, the measured cluster multiplicity is very similar for the 100 keV beam and for the low energy beam, which already indicates that the backscattering fraction at 4 or 6~keV is likely  to be very small (assuming that it is vanishingly small at 100~keV).

\begin{figure*}[tb]
  \centering
    \centering
    \includegraphics[width=1.\linewidth]{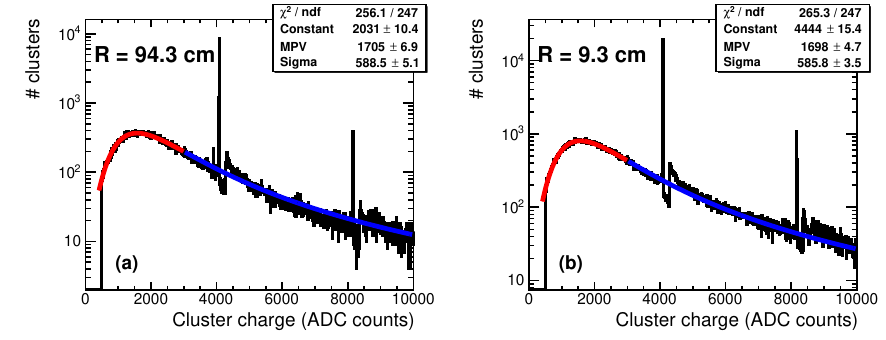}
    \caption{Distribution of the charge of the selected clusters seen in the analysis of the 6~keV data for two example positions of the CMOS sensor, $R = 94.3$~cm (a) and $R = 9.3$~cm (b). In each panel, the overlaid curve shows the result of a fit of a Landau function to the measured distribution. The fit is performed in the range between 500 and 3000 ADC counts (red curve), and the extension of the resulting function to the full range is shown in blue.   }
    \label{fig:cmos_cluster_charge}
\end{figure*}

\paragraph{ADC linearity} As seen in Fig.~\ref{fig:mcp2-Nclusters-raw-data}, the number of clusters in the CMOS sensor is very large, above 20000, when the sensor is very close to the beam line. It is important to check that the ADC is not saturating even in this high density regime: saturation effects would indeed flatten the rise of the cluster multiplicity with $1 / R^2$, while this flattening is precisely our handle on the backscattering fraction. The linear behaviour of the ADC has been checked in a run taken while ELENA was using a non standard extraction scheme which resulted in  ``mega-bunches'' being delivered to GBAR, with an intensity larger by a factor of $3$ than that of usual bunches. The beam energy was 100~keV, and the CMOS sensor was at $R = 9.3$~cm. 
The number of clusters measured in events taken with nominal bunches and with mega-bunches was seen to scale very well with the bunch intensity measured by the ELENA induction monitors~\cite{Marqversen:2022igq}, which shows that there is no significant saturation effect even in this high density regime~\cite{GBARNote}.

\paragraph{Clustering algorithm}
It has been checked that the distribution of the charge of the selected clusters does not depend on the  position of the CMOS sensor. This is illustrated in Fig.~\ref{fig:cmos_cluster_charge}, which shows the cluster charge distributions seen in the runs taken with 6~keV antiprotons impinging on  \MCPTwo, 
for $R = 9.3$~cm and $R = 94.3$~cm.
The saturation peak of the 12 bits ADC (4096 ADC counts) is clearly visible, and the peak at twice this value corresponds to clusters made of two pixels which both saturate. The structures seen after the  peaks are due to the requirement that only pixels with a charge larger than 200~ADC counts are considered in the clustering algorithm. A Landau function is fitted to each distribution in the range between 500 and 3000 ADC counts (red curves in the plots of Fig.~\ref{fig:cmos_cluster_charge}). All fits are good as shown by the corresponding $\chi^2$ values, and the extension of the fitted Landau functions to the full range (blue curves) is seen to provide a good description of the measured distributions. The parameters of the Landau functions (most probable value and width) are given in each panel of Fig.~\ref{fig:cmos_cluster_charge}. Within the statistical uncertainties of the fits, these parameters do not depend on the  position of the CMOS sensor. This observation shows that similar clusters are seen at all distances, and that the clustering algorithm is not merging or splitting clusters when the sensor is at small distances from the beam line. Hence, the clustering does not bias the $R$-dependences seen in Fig.~\ref{fig:mcp2-Nclusters-raw-data}. The fits also show that the requirement that the cluster charge exceed 500 ADC counts has a very high efficiency.

\section{Fit of the backscattering fraction of antiprotons on \MCPTwo} 
\label{sec:backscattering-results}

The linear fits shown in Fig.~\ref{fig:mcp2-Nclusters-raw-data}, performed to the large $R$ measurements ($R \geq 39$~cm), show an offset  
not observed in the simulations 
(Fig.~\ref{fig:G4_final_templates_4keV}),
due to beam induced noise. This offset is indeed  larger than the ``dark'' noise measured in the sensor when the ELENA beam is 
blocked.
Dedicated background runs, during which the beam was passing through the experiment but where \MCPTwo had been taken out of the line, confirmed a non-zero number of clusters in the sensor, after subtracting the pedestals measured with no beam. These clusters arise from annihilations elsewhere along the line.
This  offset is subtracted from the measurements shown in Fig.~\ref{fig:mcp2-Nclusters-raw-data}, and a fit of the GEANT4 predictions is made to these subtracted data. \\

\subsection{Fit model}
\label{sec:fitmodel}

The simulations described in section~\ref{sec:G4_fullmodel} provide a prediction for the number of clusters measured in the CMOS sensor as a function of the  distance between the sensor and the centre of the MCP, for $E=4$~keV or $E=6$~keV:
\begin{eqnarray*}
\begin{array}{lllllcll}
p_E (R)   &=   & N_E  \quad  \times  & [   & \,  & (1 - f_{\rm{back},E })  & \times  \quad P_{\rm{annihil},E}( R, \delta )  & \\
          &    &                     & \,  & +    & f_{\rm{back},E}         & \times \quad P_{\rm{back},E} ( R, \delta )  & ]
\end{array}
\label{eq:prediction_lowE}
\end{eqnarray*}
where $f_{\rm{back},E}$ is the backscattering fraction on \MCPTwo at the energy considered, and $N_E$ is an overall normalisation factor. In the equation above, $\delta$ denotes a possible longitudinal shift of  the CMOS sensor with respect to its nominal position,
used in the GEANT4 model. 
This shift is given by the sum of:
\begin{itemize}
    \item a potential constant shift $l_s$, independent of the radial position of the sensor, resulting from the uncertainty on the longitudinal distance between the centre of the sensor and the MCP (measured to be $\sim 8.5$~mm when the sensor is at $R = 9.3$~cm); 
    \item and a component that depends on the  radial position of the sensor, resulting from a potentially non-zero angle $\alpha$ between the direction of the rail along which the sensor was moved, and the plane orthogonal to the beam direction.
\end{itemize}

\begin{figure*}[htbp]
  \centering
    \centering
    \begin{tabular}{cc}
    \includegraphics[width=0.45\linewidth]{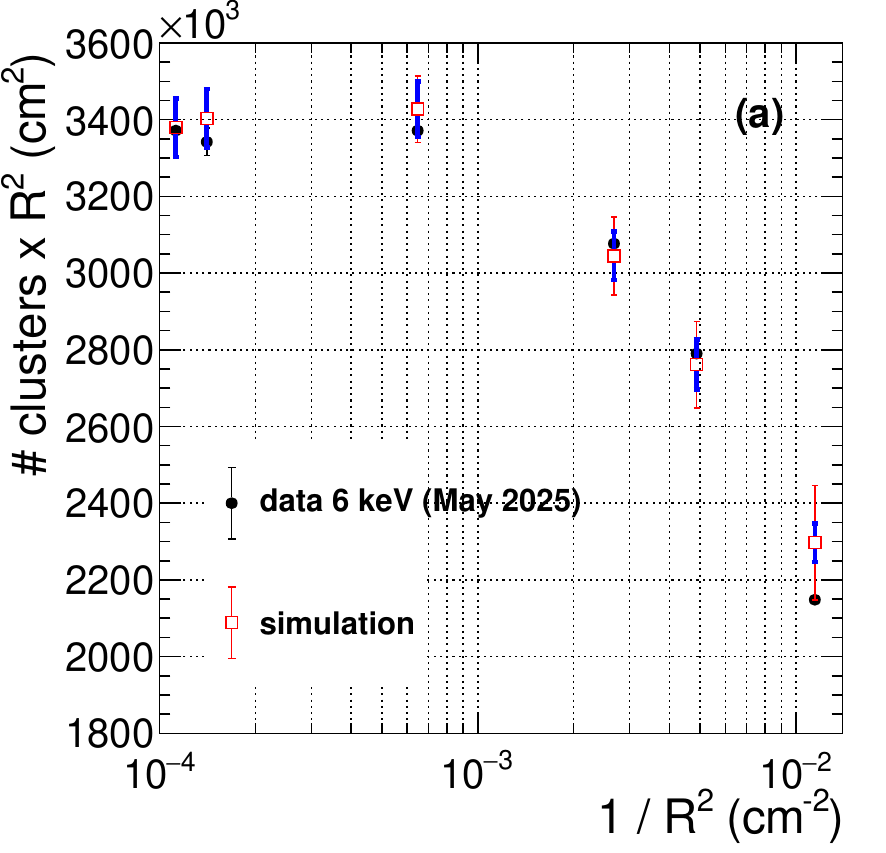} &    
    \includegraphics[width=0.45\linewidth]{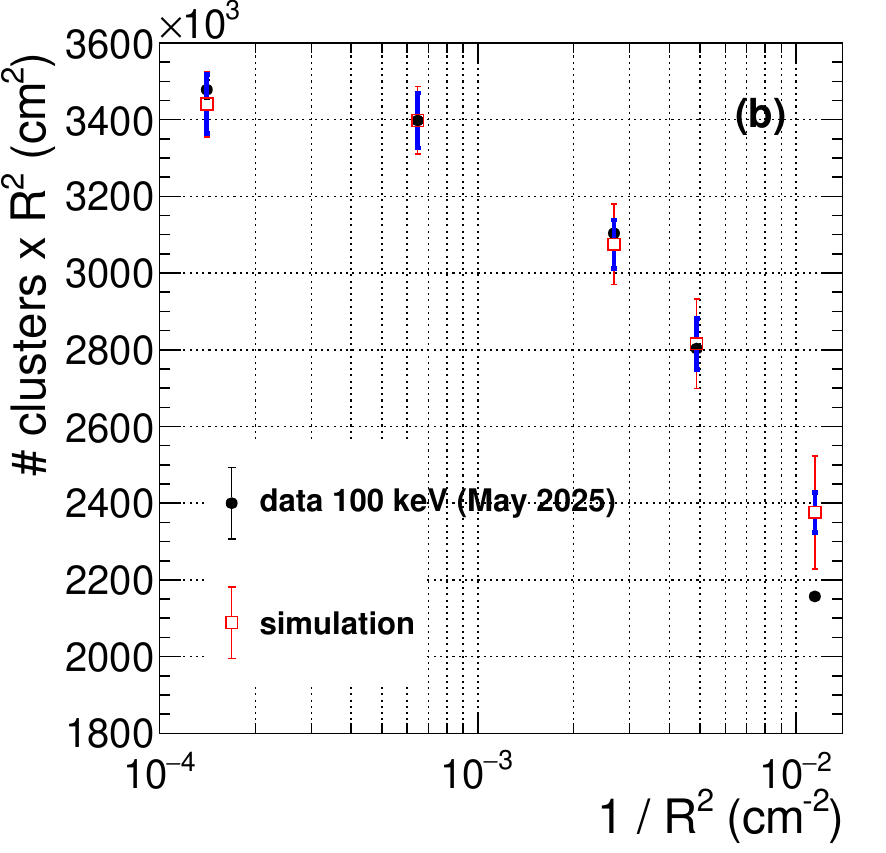}  \\
    \includegraphics[width=0.45\linewidth]{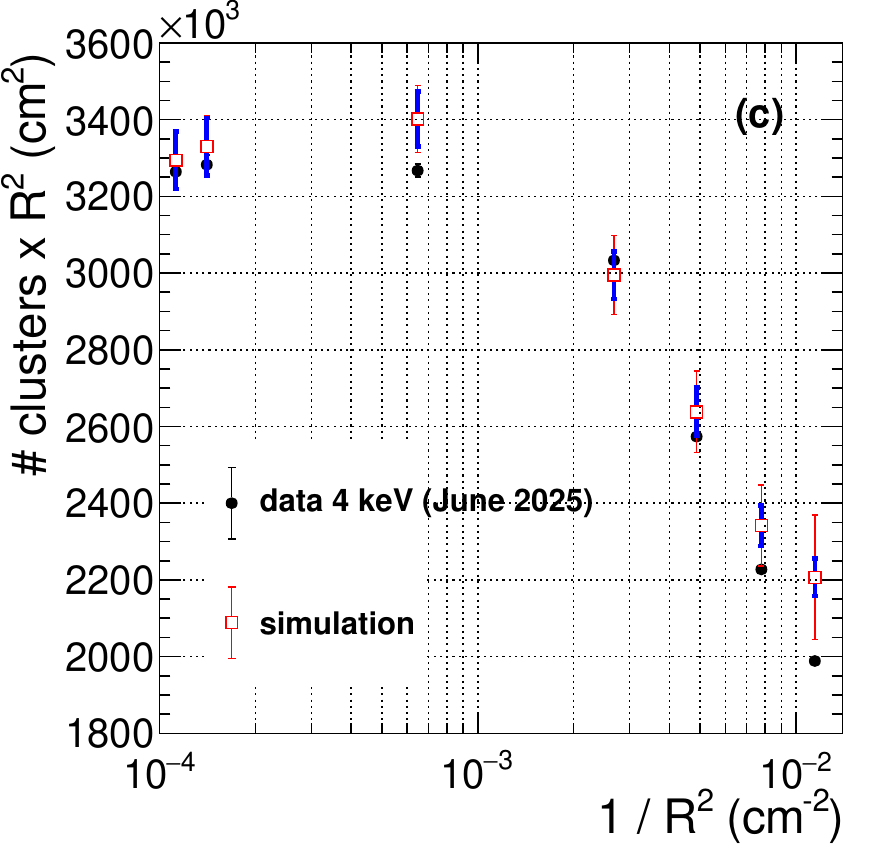} &
    \includegraphics[width=0.45\linewidth]{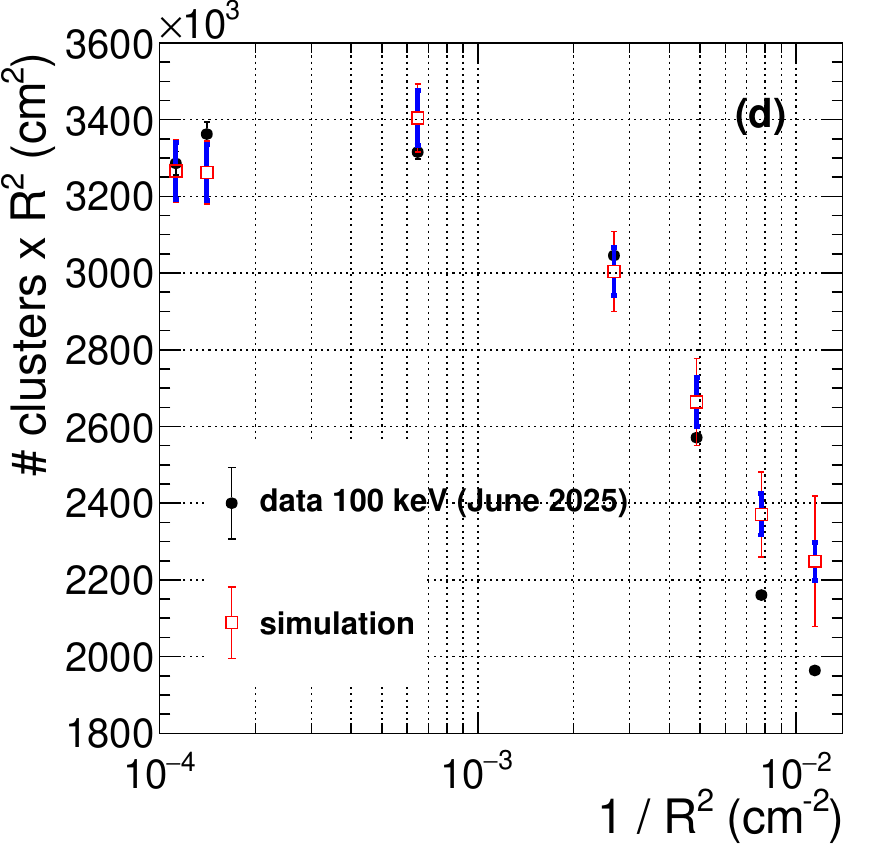}     
    \end{tabular}
    \caption{ 
    (a) and (b): Number of clusters measured in the CMOS sensor (data from May 2025) as a function of the inverse of the squared  distance between the sensor and the centre of the MCP  (full black circles), and resulting from the fit of the GEANT4 predictions as described in the text (open red squares). The number of clusters is scaled by $R^2$ in order to show better the points at large distances. The thick blue line around each red square represents the uncorrelated uncertainty of the prediction, while the thin red line represents the full uncertainty. The left (right) plot corresponds to a kinetic energy of 6~keV (100~keV) for the incoming antiprotons.  (c) and (d): idem, for the 4~keV and 100~keV data from June 2025.
    }
    \label{fig:fitresult}
\end{figure*}

The backscattering fraction of 100~keV antiprotons is assumed to be negligibly small (as predicted by GE\-ANT4, irrespective of the MCP coating and of the modelling). Consequently, the number of clusters that should be seen in the CMOS sensor at 100~keV is:
$$ p_{100}(R) =  N_{100}  \times   P_{\rm{annihil}, 100}( R, \delta )  \quad $$
where $N_{100}$ is another normalisation factor.
Within a given data taking campaign, the parameters $l_s$ and $\tan \alpha$ that define $\delta$ are the same for the low energy and the 100~keV measurements. \\

Two combined fits are run independently, one to the 6~keV and 100~keV data taken in May 2025, the other to the 4~keV and 100~keV data taken in June 2025. Each fit has five free parameters: the backscattering fraction at 6 or 4~keV; the normalisations $N_{100}$ and $N_E$ of the 100~keV and of the low energy measurements; the longitudinal shift $l_s$; and the tangent of the angle of the rail, $\tan \alpha$.
Each combined fit is made by minimizing a $\chi^2$ function between the numbers of clusters measured in the CMOS detector and the corresponding predictions.
The latter are affected by the following systematic uncertainties:
\begin{enumerate}
    \item An uncertainty of 2~mm is set on the radial distance between the sensor and the beam line, that is:
    \begin{itemize}
        \item uncorrelated between the various radial positions at which measurements were made;
        \item at a given value of $R$, fully correlated between the measurement at 100~keV and the measurement at the low energy.
    \end{itemize} 
    \item The simulation accounts for the actual position of the beam when it hits \MCPTwo.
    A shift of the transverse position of the beam spot with respect to what has been used in the simulation,
    which could result from an off-centred imaging of the beam,
    would translate into a shift on the radial distance between the sensor and the beam line, identical for each energy and (to first order) for each value of $R$. This is accounted for by setting another systematic uncertainty of 2~mm on the radial distances, that is fully correlated between all points.
    \item An additional uncorrelated  uncertainty, of $2 \%$ in relative, has been set to all predictions. It aims at accounting for residual mis-modellings, coming, for example, from a non perfect description of the material in the vicinity of the MCP.  
\end{enumerate}

The combined fit to the June 2025 data (4~keV and 100~keV) uses 14 measurements, resulting in 9 degrees of freedom. In the data taken in May 2025, no measurement was made at $R = 11.3$~cm and the point at $R=94.3$~cm was not measured at 100~keV, resulting in 11 data points and 6 degrees of freedom in the fit. The $\chi^2$ function is minimised using the MINUIT package~\cite{James:2004xla}.

\subsection{Results of the fit to the 4 keV + 100 keV data}
\label{sec:fit_results_4-100}

The combined fit to the 4~keV and 100~keV data taken in June 2025 results in a good minimum $\chi^2$ , $\chi^2_0 =10.9$.  
The uncertainty on the radial distance of the sensor to the beam line, correlated between the 100~keV and the 4~keV points, has a significant impact on the quality of the fit: neglecting it increases the $\chi^2$ to $\sim 17$. Neglecting the uncertainty on the beam spot position has a smaller impact, the $\chi^2$ increasing to $\sim 13.5$. Ignoring the $2\%$ uncorrelated uncertainty increases the $\chi^2$ to $\sim 23$. \\

At the minimum of the $\chi^2$, the backscattering fraction at 4~keV is zero, the longitudinal shift $l_s$ is $-2.2$~mm, and the angle $\alpha$ is $11$~mrad. Figure~\ref{fig:fitresult}c-d  shows how the combined fit describes the  measurements.
The overall description is good, even though the fit slightly overestimates the number of clusters measured at small distances, both at 100~keV and at 4~keV. The uncertainty on the fit predictions at small values of $R$ is completely dominated by its correlated component as shown by the thin error bars in Fig.~\ref{fig:fitresult}c-d, such that the four ``pulls" corresponding to the points at $R=9.3$~cm and $R=11.3$~cm at the two energies are strongly correlated.  \\

\begin{figure}[tbhp]
  \centering
    \centering   
    \includegraphics[width=0.9\linewidth]{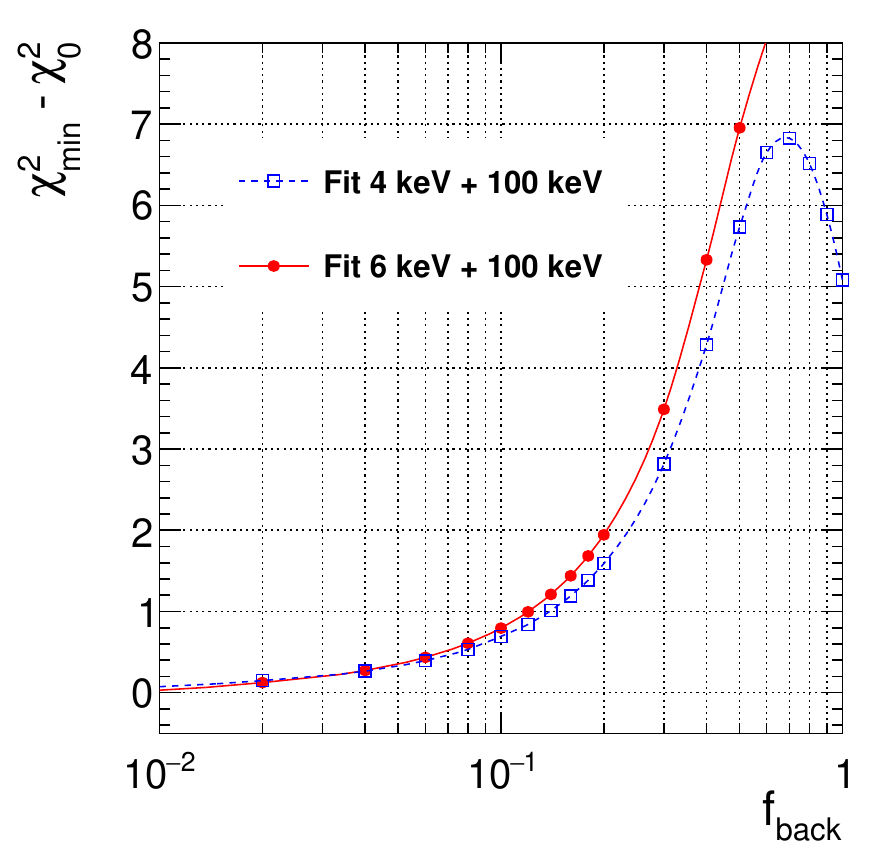}
    \caption{ 
    Minimum of the $\chi^2$ function that is minimised in the combined fit to the 6~keV and 100~keV data from May 2025 (red closed circles, full curve), or in the fit to the 4~keV and 100~keV data from June 2025 (blue open squares, dashed curve) as a function of the backscattering fraction at 6~keV (4~keV). The absolute minimum $\chi^2_0$, obtained at $f_{\rm{back}} = 0$ in both cases, is subtracted from each point.
     }
    \label{fig:scan_chi2}
\end{figure}

The uncertainties on the fitted parameters have been derived from a full scan of the $\chi^2$ function (the $\chi^2$ being minimised over the remaining free parameters when scanning over the parameter of interest). The angle $\alpha$ is seen to be determined by the fit with an uncertainty of $5$~mrad, while $l_s$ is determined within $1$~mm. The resulting scan for the backscattering fraction is shown in Fig.~\ref{fig:scan_chi2} (dashed curve).  The range of values of the backscattering fraction that is consistent with the criterion $\Delta \chi^2 \leq 1$, where $\Delta \chi^2 = \chi^2_{min} - \chi^2_0$, is $f_{\rm{back}} \lesssim 14\%$.   \\

In order to assess the residual model dependence resulting from potential re-scattering of the backscattered antiprotons on the beam pipe (or on the vacuum cross), the probabilities $P_{\rm{back}}(R)$ and $P_{\rm{annihil}}(R)$ have been determined in an alternative GEANT4 simulation, where the $\theta_{lim}$ parameter of the WentzelVI model has been set to 2.5~radians (the range cut being kept at 0.2). With respect to the nominal settings, the re-scattering probability of backscattered antiprotons on the pipe or on the cross is then decreased by a factor of two. The probability $P_{\rm{back}}(R)$ is much less affected, changing by at most $9 \%$ (as expected, the probability $P_{\rm{annihil}}(R)$ is unchanged). Repeating the fit with these alternative  probabilities leads to very similar results, in particular the backscattering fraction at the minimum of the $\chi^2$ function remains zero. The quality of the fit is only slightly degraded, with a minimum $\chi^2$ of 11.9. \\

The fit has also been repeated with the probabilities $P_{\rm{annihil}}(R)$ and $P_{\rm{back}}(R)$ resulting from an alternative GEANT4 simulation, in which the  {\tt {BERT}} physics list is used in place of {\tt {INCL}}. Despite differences between the probabilities predicted by the two models that can reach several per-cents, the two fits are very similar. The fit based on {\tt{BERT}} also results in a vanishing backscattering fraction and in very similar values as the nominal fit for the parameters $l_s$ and $\tan \alpha$. The $\chi^2$ at the minimum, 9.7, is slightly lower than for the nominal fit, and the uncertainty bound on $f_{\rm{back}}$ is similar, $f_{\rm{back}} \lesssim 13\%$ at $68 \%$ confidence level.

\subsection{Results of the fit to the 6 keV + 100 keV data}

The combined fit to the 6~keV and 100~keV data taken in May 2025 results in a good $\chi^2$ of 4.7 for 6 degrees of freedom. The minimum of the $\chi^2$ is reached for a vanishing backscattering fraction at 6~keV, a longitudinal shift $l_s$ of $+1.1$~mm and an angle $\alpha$ of $1$~mrad. The parameters $l_s$ and $\tan \alpha$ do not have to be  identical to those determined from the fit to the June 2025 data, since the setup had been moved between the two data taking periods.
The overall description of the measurements is good, as shown by Fig.~\ref{fig:fitresult}a-b.
As shown in Fig.~\ref{fig:scan_chi2} (full curve), the range of values of the backscattering fraction at 6~keV that is preferred by this fit is $f_{\rm{back}} \lesssim 11.5\%$ at $68 \%$ confidence level. 
An alternative fit to predictions obtained with $\theta_{lim}$ set to 2.5 radians also results in a vanishing backscattering fraction, with a very similar $\chi^2$ at the minimum, 4.5. 
The results of an alternative fit to  predictions obtained using the {\tt {BERT}} physics list in place of {\tt{INCL}} are very similar to those obtained from the nominal fit.

\section{Determination of the flux of antiprotons for antihydrogen production}

\begin{figure*}[tbh]
  \centering 
  \begin{tabular}{cc}
    \includegraphics[width=0.4\linewidth]{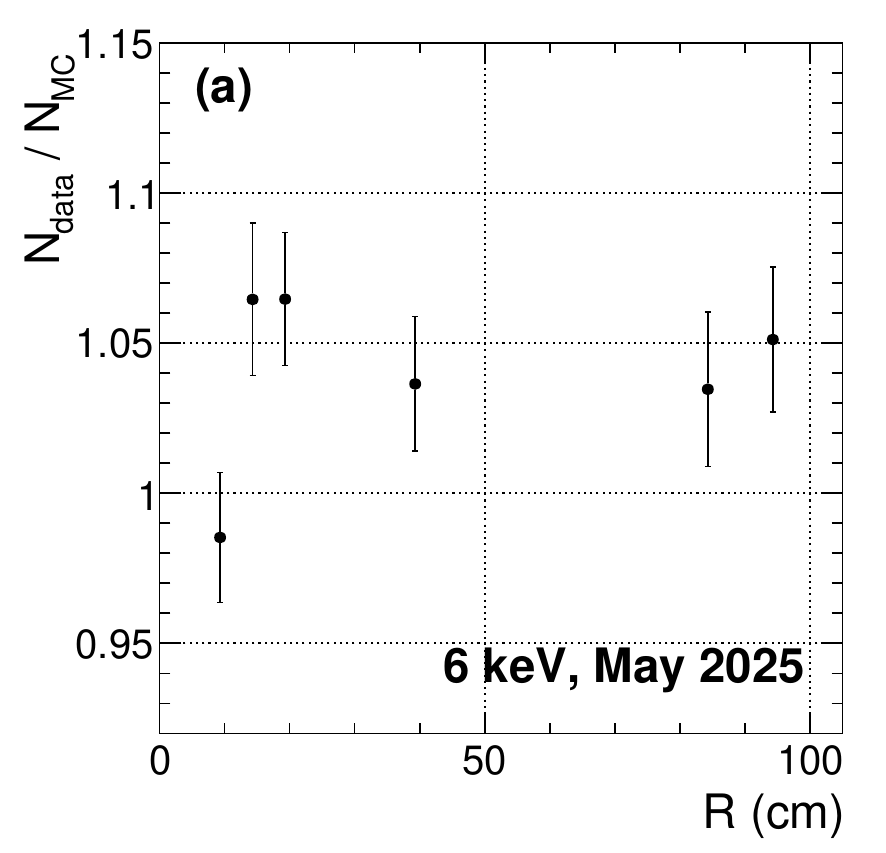}     &
    \includegraphics[width=0.4\linewidth]{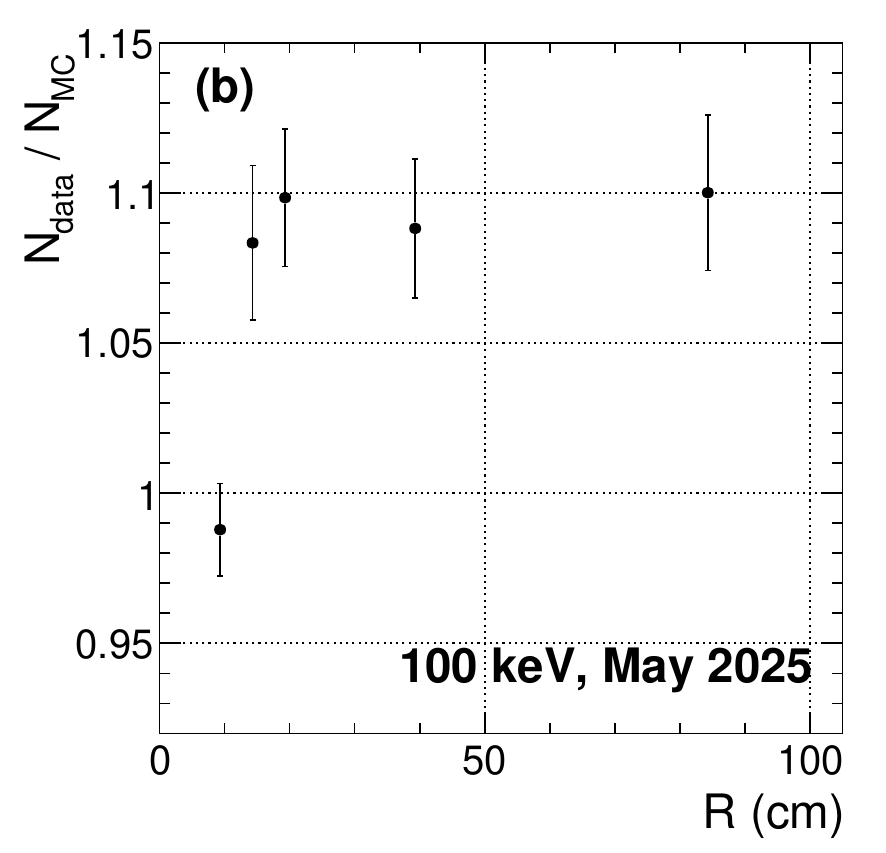} \\
    \includegraphics[width=0.4\linewidth]{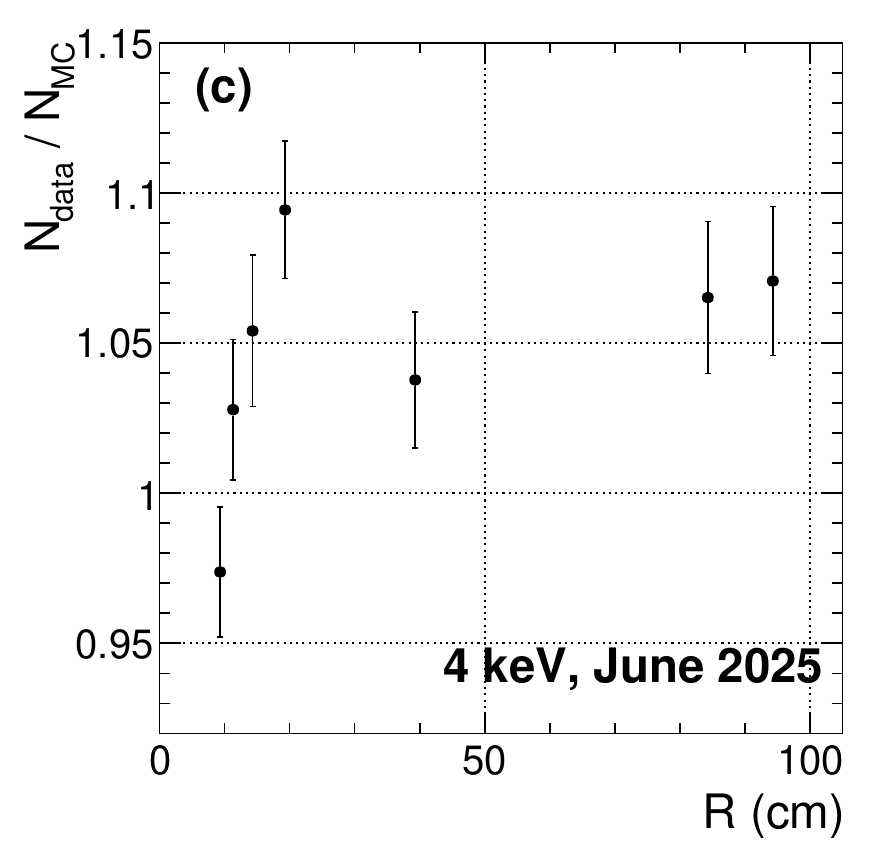} &
    \includegraphics[width=0.4\linewidth]{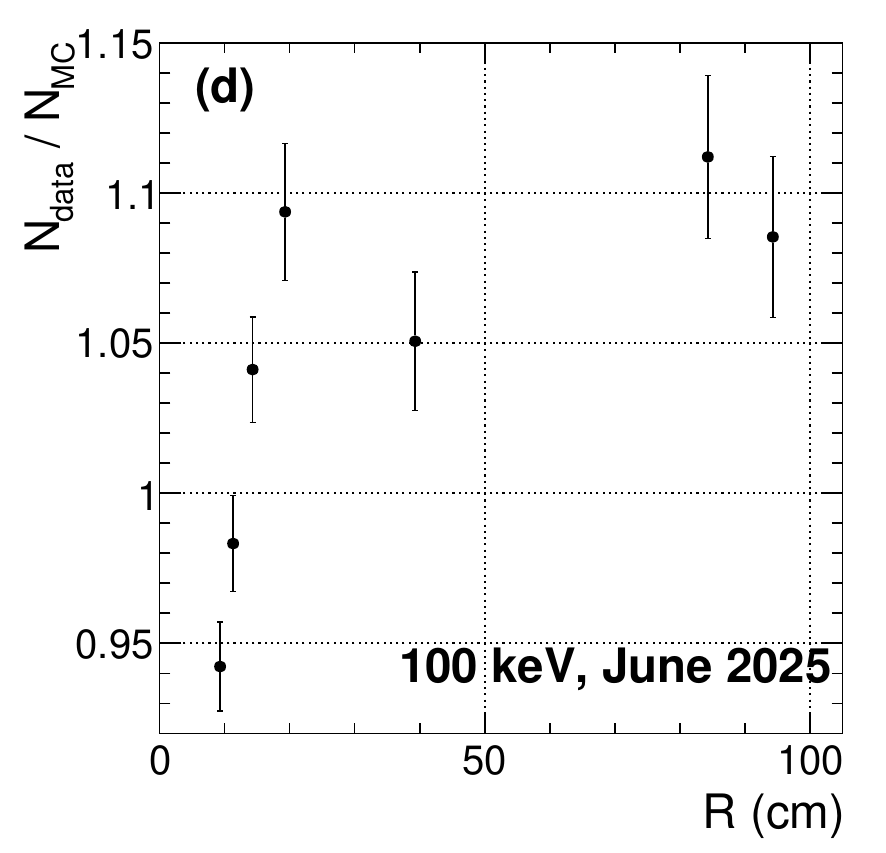}
    \end{tabular}
   \caption{ 
   Ratio of the number of clusters measured in the CMOS sensor to the one predicted from the simulation, for the data taken with the \MCPTwo setup. The prediction uses the incoming antiproton intensities measured by ELENA. The error bars include the statistical uncertainty of the measured number of clusters and the uncorrelated uncertainty of the predicted number of clusters per annihilating antiproton. 
   }
    \label{fig:validation_normalisation_MCP2}
\end{figure*}

\subsection{Calibration of the simulation  using the \MCPTwo data}

The previous section has shown that the simulation correctly describes the $R-$dependence of the number of clusters observed in the CMOS sensor. The focus of this paragraph is the absolute scale of these GEANT4 predictions.

For the data described previously, 
the intensity of antiprotons impinging on \MCPTwo is equal to the intensity of the beam delivered to GBAR by ELENA. Since the latter intensity is measured, shot by shot, by ELENA monitors (with a systematic uncertainty of about $4\%$, correlated between all shots), 
the number of clusters predicted from the simulation is $N_{MC} = P \times I_{EL}$, where $I_{EL}$ denotes the ELENA intensity and $P$ the number of clusters predicted per incoming antiproton that annihilates on the MCP.
Hence, the comparison of the number of clusters measured in the CMOS sensor, $N_{data}$, with the prediction $N_{MC}$, provides a calibration of the simulation. 

This comparison can be seen in Fig.~\ref{fig:validation_normalisation_MCP2}. 
The predictions account for the values of $l_s$ and $\tan \alpha$ determined by the fits, but differ from the predictions shown in Fig.~\ref{fig:fitresult} since the overall normalisation of the predicted number of clusters is now fixed by the ELENA intensity ($P \times I_{EL}$), instead of being a free parameter. 
For distances larger than $11$~cm, the ratio $ N_{data} / N_{MC}$, within a given dataset, does not depend on the distance\footnote{At the smallest distances, the prediction for the number of clusters is heavily affected by correlated systematic uncertainties which can reach $8\%$, as was shown in Fig.~\ref{fig:fitresult}.}. 
It is on average higher than unity by $7.4 \%$. Taking into account the uncertainty of the ELENA intensities, this value could be explained by a $6\%$ uncertainty of the normalisation of the Monte-Carlo predictions. Following the approach used in Ref.~\cite{CR_paper}, the choice is made to scale up by $7.4 \%$ the GEANT4 predictions that will be shown in the next paragraph. 

\begin{figure*}[tbhp]
  \centering
    \centering
    \begin{tabular}{cc}
    \reflectbox{\includegraphics[width=0.4\textwidth]{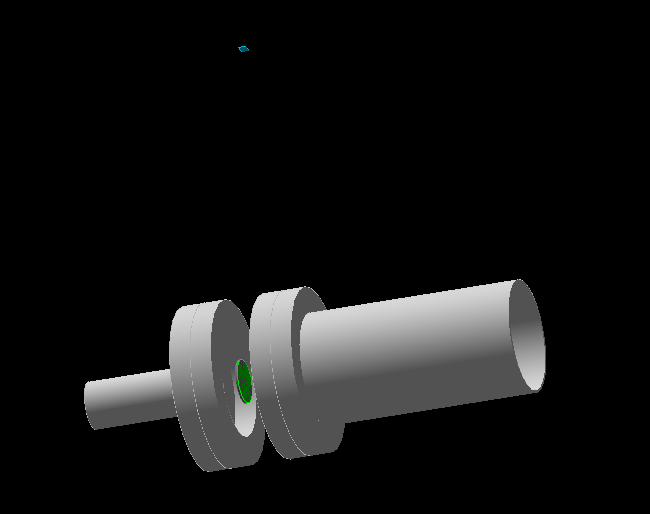}} & 
    \includegraphics[width=0.4\linewidth]{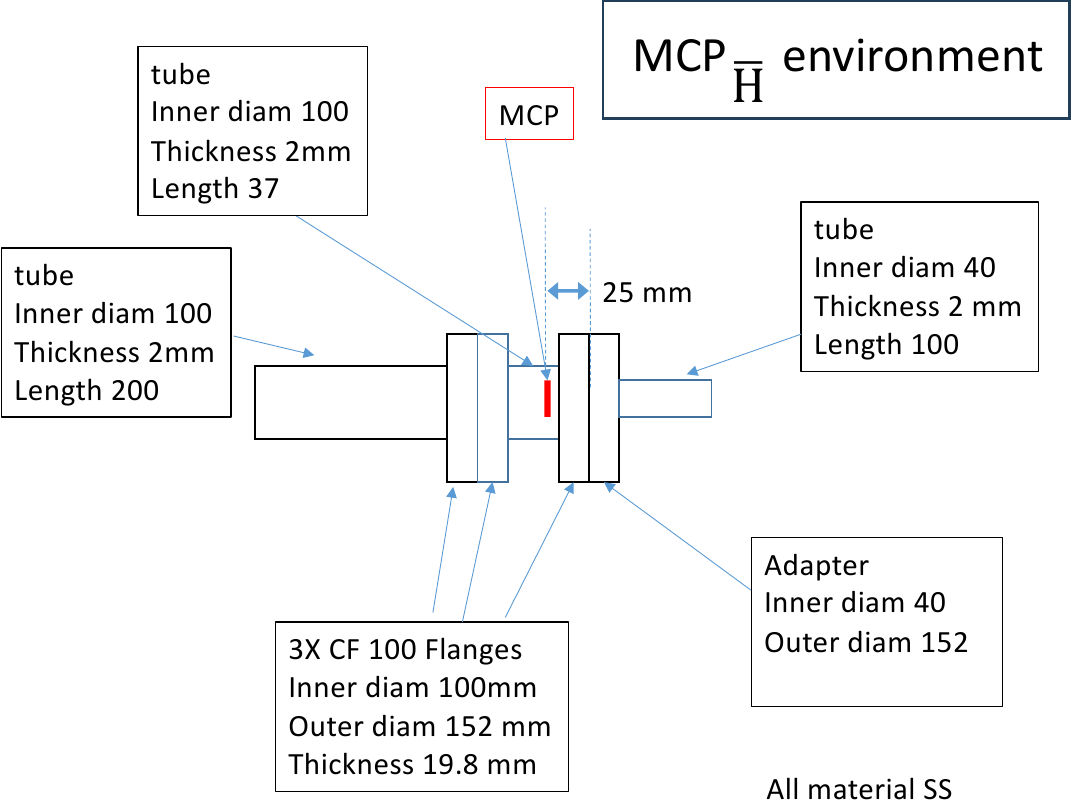}
    \end{tabular}
    \caption{Left: Visualisation of the simulation geometry around \MCPFive. For visualisation 
    purposes, the beam pipe around the MCP was removed to expose the internal components. 
    The small light blue cuboid above is the CMOS detector. Right: Schematic drawing of geometry components near \MCPFive. The CMOS detector is omitted in this drawing. }
    \label{fig:MCP5_BL}
\end{figure*}

\subsection{Calibration of the flux impinging on \MCPFive }

In GBAR, the number of antiprotons that participate in antihydrogen production is monitored regularly during special ``pass-through'' runs where the antiprotons traverse the reaction chamber and annihilate downstream on \MCPFive. The CMOS sensor, placed above \MCPFive, detects the products of the antiproton annihilations. 
The ratio between the  number of clusters measured in the CMOS detector and  the number of clusters predicted per incoming antiproton that annihilates on the MCP provides an estimation of the number of incident antiprotons~\cite{CR_paper}.

A simulation of antiproton annihilation on \MCPFive has been performed using the procedure described above (see Sections~\ref{sec:modelling} and~\ref{sec:G4_EMlist}). 
Antiprotons with a kinetic energy of 6~keV are directed toward the MCP surface along the nominal beam axis.
The simulated beam line geometry around \MCPFive includes beam pipes, vacuum flanges, and the MCP 
assembly, all modelled in stainless steel to match the actual experimental configuration 
(see Figure~\ref{fig:MCP5_BL}).
The \MCPFive substrate is modelled as a plain disk of lead-glass as done for \MCPTwo (see Section~\ref{sec:G4_fullmodel}), of radius 2~cm and of thickness 2~mm. 
The 20 nm thick surface coating differs: 
while \MCPTwo uses chromium, the manufacturer of \MCPFive specified a NiCr coating without 
providing the exact alloy composition. A typical  alloy composition was adopted 
(80\%~Ni, 20\%~Cr).
Unlike the \MCPTwo configuration, no mounting rings or other structural elements are positioned in 
front of \MCPFive.
The CMOS position is fixed at 34~cm above the MCP centre, corresponding to its nominal location during the 2024 data run. \\

\begin{table}[htbp]
\caption{Number of clusters in the CMOS sensor, resulting from a simulation of antiprotons impinging on \MCPFive. The numbers are given for all antiprotons, and separately for those that annihilated on \MCPFive and those that backscattered.}
\label{tab:MCP5_result}
\begin{tabular}{ lll }
\hline\noalign{\smallskip}
 & Nb of $\overline{p}$ & Nb of clusters   \\
 \noalign{\smallskip}\hline\noalign{\smallskip}
Total & 100000000  & 20111   \\  
Annihilated on \MCPFive & 59856979   & 13428  \\ 
Backscattered& 40009296  & 6660  \\ 
\noalign{\smallskip}\hline
\end{tabular}
\end{table}

The rightmost column in the first line of Table~\ref{tab:MCP5_result} shows the predicted number of clusters in the CMOS sensor when 100 million antiprotons impinging on \MCPFive are simulated. 
In the second and third lines, this number is given separately for the antiprotons that annihilate on \MCPFive and for those that backscatter (about $40 \%$ of the sample, consistent with Fig.~\ref{fig:G4_backscattering_vs_Ekin_various_materials}). Similar to what was seen in the simulations on \MCPTwo, the backscattered antiprotons have a lower probability of producing clusters in the CMOS sensor compared to antiprotons that annihilate on the MCP: for antiprotons that annihilate on \MCPFive, this probability amounts to $2.24 \times 10^{-4}$,  while it is $1.66 \times 10^{-4}$  for backscattered antiprotons. According to what was shown in the previous paragraph, these probabilities need to be scaled up by $7.4 \%$.

The previous section has shown no evidence for back\-scat\-te\-ring off \MCPTwo.
Under the reasonable assumption that the backscattering fraction off the \MCPFive surface is equally negligible, the number of antiprotons that hit \MCPFive, when running in pass-through mode, is obtained by multiplying the number of clusters observed in the CMOS detector by $ 1/ ( 1.074 \times 2.24 \times 10^{-4}) = 4156$.

Allowing for non-vanishing and uncorrelated back\-scat\-te\-ring fractions on \MCPTwo and \MCPFive, respectively $f_1$ and $f_2$, the GEANT4 probability to see clusters in the CMOS sensor when antiprotons impinge on \MCPFive would scale as $ ( 1 - f_2 ) \times 2.24 \times 10^{-4} + f_2 \times 1.66 \times 10^{-4}$, while the factor by which the GEANT4 predictions should be scaled, according to the comparison of the prediction with the 6 keV measurement on \MCPTwo when the CMOS sensor is at a distance $R$ from the MCP centre, would scale as the inverse of 
$  ( 1 - f_1 ) \times P_{\rm{annihil}, 6} ( R )  + f_1 \times P_{\rm{back},6}( R )  $.
The quantity by which the CMOS calibration factor given above would need to be scaled has been determined for  $f_1$ and $f_2$ varying independently from each other; a conservative variation range, between zero and $20\%$, has been chosen.
As expected, this quantity is very close to unity when $f_1 = f_2$.
When $f_1$ is below $14 \%$, it varies from unity by at most $5 \%$, with a negligible dependence on the value of $R$ chosen to determine the scale factor of the simulation; this maximal variation is obtained for a maximal asymmetry between $f_1 $ and $f_2$, namely $f_2 = 20 \%$ and $f_1 = 0$.
Consequently, a systematic uncertainty of $5 \%$ is set on the CMOS calibration factor, to account for possibly different backscattering fractions on \MCPTwo and on \MCPFive.

This calibration factor is affected by several other sources of systematic uncertainties,
as given in Ref.~\cite{CR_paper}.
An uncertainty of $0.9\%$ comes from the finite Monte-Carlo statistics entering in Tab.~\ref{tab:MCP5_result}.
The factor used to rescale the GEANT4 predictions is affected by a $4\%$ uncertainty on the ELENA intensities, and an additional uncertainty of $4\%$ is assigned based on the fluctuations of the ratios shown in Fig.~\ref{fig:validation_normalisation_MCP2}.
An additional systematic uncertainty arises from the position of the antiproton beam on \MCPFive, which, in 2024, could be off-centred vertically by typically 5~mm. This radial
offset modifies the distance between the annihilation point and the CMOS detector 
from the nominal 34~cm to $34 \pm 0.5$~cm, resulting in a relative variation of about $3\%$ of the 
geometric acceptance of the CMOS sensor,
i.e. in an additional systematic uncertainty of $3\%$ on 
the derived number of incident antiprotons.
This number is also affected by a $2\%$ uncertainty resulting from an uncertainty of $\pm 1$~mm on the longitudinal position of the CMOS sensor with respect to the centre of  \MCPFive.
Consequently, the overall systematic uncertainty on the 
number of  antiprotons that impinge on \MCPFive 
is estimated to be $\sim 8.4\%$. \\

The calibration factor obtained here is $9\%$ lower than the one given in Ref.~\cite{CR_paper}. 
Out of this difference, about $3 \%$ 
appear to be due to a slightly different simulation. The residual $6\%$ difference is well within the systematic uncertainties that are uncorrelated between the two determinations, which rely on independent \MCPTwo measurements.

\section{Conclusions}

The probability that low energy antiprotons backscatter from the surface of the MCPs used in the GBAR experiment has been determined in a dedicated measurement campaign.
A CMOS sensor placed at different distances from the beam axis was used to detect the products of 
the annihilations of 4, 6 and 100~keV antiprotons impinging on a MCP. The dependence of the signal
measured in the sensor with this distance provides a handle on the fraction of antiprotons that
backscatter. The measurements show no evidence for backscattering down to 4 keV, and an upper bound
of $14 \%$ has been set on this fraction at $68 \%$ confidence level. The systematic uncertainty on the number of antiprotons that
participate in antihydrogen production in the GBAR experiment, that results from uncertainties
on the backscattering fraction, has also been derived.

\begin{acknowledgements}
We thank L. Ponce and the AD/ELENA team as well as F. Butin and the CERN EN team for
their fruitful collaboration.  
This work is supported by: JSPS KAKENHI Grant-in-Aid for Scientific Research A
20H00150 and Fostering Joint International Research A 20KK0305 (Japan), 
the Action Thématique GRAM of CNRS/INSU with INP and IN2P3 co-funded by CNES (France),
SPHINX ANR-22-CE31-0019 (France), ESPRIT ANR-22-CE30-0028-01 (France), BESCOOL ANR/DFG ANR-13-ISO4-0002-01 (France/Germany), the Swiss National Science Foundation (Switzerland) grants
197346, 216673 and 232699 and ETH Zurich (Switzerland) grant ETH-46 17-1, the Swedish
Research Council (VR) grants 2017-03822, 2021-04005 and 2025-04378, the German cluster of excellence PRISMA, and the following grants from Korea:
IBSR016-Y1, IBS-R016-D1, UBSI Research Fund (No. 1.220116.01) of UNIST, POSTECH Initial
Settlement Support Fund, NRF-2016R1A5A1013277, NRF-RS-2022-00143178,
NRF-2021R1A2C3010989, and NRF-2016R1A6A3A11932936. The GBAR collaboration is an
International Research Network, supported by CNRS, France.

\end{acknowledgements}


\bibliographystyle{spphys.bst}
\bibliography{bibfile}

\end{document}